\documentclass[twocolumn]{openjournal}
\usepackage{graphicx}   
\usepackage{amsmath}    
\usepackage{amssymb,amsfonts}    
\usepackage{orcidlink}
\usepackage{verbatim}
\usepackage{hyperref}
\usepackage{xcolor}
\usepackage{longtable}
\usepackage{booktabs}
\usepackage{float} 

\shorttitle{Parameters of the binary Nandou=Phi Sagittarii}
\shortauthors{Idel Waisberg \& Boaz Katz}

\begin{document}

\title{Brewing another binary in the Teapot:\\Nandou=Phi Sagittarii} 

\footnotetext[]{Based on observations collected at the European Southern Observatory, Chile, Program IDs 099.D-2020(A), 113.26HF.001}

\newcommand{\weizmann}{Department of Particle Physics and Astrophysics, Weizmann Institute of Science, Rehovot 76100, Israel}

\email{email: idelwaisberg@gmail.com}

\author{\vspace{-1.2cm}Idel Waisberg\,\orcidlink{0000-0003-0304-743X}$^{1}$ \& Boaz Katz\orcidlink{0000-0003-0584-2920}$^{2}$}

\affiliation{$^1$Independent researcher, Lambda Ophiuchi Ltda}
\affiliation{$^2$\weizmann}

\begin{abstract}
The iconic Teapot asterism in Sagittarius is particularly rich in intermediate-mass stars in multiple systems. Here we combine VLTI/GRAVITY, archival spectroscopic observations, past radial velocity measurements and photometry to solve for the orbital and stellar parameters of the binary Nandou=Phi Sagittarii. It has a period $P=238.844\pm0.011$ days, a semi-major axis $a=1.32\pm0.10\text{ au}$ and eccentricity $e=0.418\pm0.004$. The precision in the dynamical masses $M_A=3.8\pm0.9 M_{\odot}$ and $M_B=1.6\pm0.4 M_{\odot}$ is currently limited by the parallax $p=12.5\pm1.0 \text{ mas}$. We find that the original Hipparcos 1997 Intermediate Astrometric Data (IAD) are consistent with the expected photocenter semi-major axis $a_{\mathrm{phot}} \sim 5 \text{ mas}$, while the IAD from the Hipparcos 2007 new reduction yield a too small $a_{\mathrm{phot}} = 1.7\pm0.3 \text{ mas}$, confirming the suspicion that a too aggressive modeling of the satellite attitude can lead to overfitting that suppresses the orbital motion. Matching the primary's temperature  $T_{\mathrm{eff,A}} = 12000\pm250 \text{ K}$ and radius $R_{\mathrm{A}} = 5.1\pm 0.4 R_{\odot}$ to stellar evolutionary models implies that it is a $M_A \simeq 4 M_{\odot}$ star near the end of the main-sequence. Nandou will eventually most likely become a symbiotic star. 
\end{abstract}


\section{Introduction}
\label{sec:introduction}

The Teapot asterism in Sagittarius (Figure \ref{fig:sky_image}) is remarkably rich in interesting stars. Out of the eight stars that make up the shape, four are early-type stars and they are all multiple systems. Three of them have been shown to be rather remarkable: 

\begin{figure*}[]
\includegraphics[width=2\columnwidth]{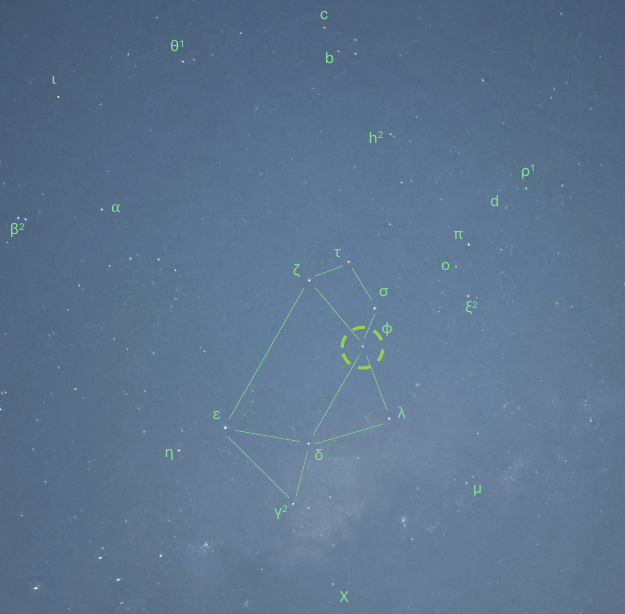}
\caption{Smartphone photo of Sagittarius as seen from São Roque, SP, Brazil on 2026-07-10 UTC 04:39. The iconic Teapot asterism is drawn. All of its four early-type stars, namely Kaus Australis ($\epsilon$ Sgr), Nunki ($\sigma$ Sgr), Ascella ($\zeta$ Sgr) and Nandou ($\phi$ Sgr), are multiple systems.}
\label{fig:sky_image}
\end{figure*}

\begin{enumerate}
\item Kaus Australis = $\epsilon$ Sgr (B9IVp) is a $3.8 M_{\odot}$ star that is rotating almost at its critical velocity \citep[the most critical rotator known in the Galaxy according to][]{Bailey2024} and is strongly suspected of being a merger product. It also has a $M\simeq1 M_{\odot}$ companion at a projected separation of about 100 au \citep{Golimowski1993}; 

\item Nunki = $\sigma$ Sgr (B2.5V) is a $6.5 M_{\odot} + 6.3 M_{\odot}$, $a=1.3 \text{ au}$ eccentric ($e\approx0.5$) binary that could eventually merge; in this case, it would be the closest core collapse progenitor to the Sun \citep{Waisberg2025,Waisberg2026b}; 

\item Ascella = $\zeta$ Sgr (A2.5Va) is a compact triple system containing a $1.8 M_{\odot} + 1.8 M_{\odot}$, $a\sim 0.3$ au binary orbiting a $2.0 M_{\odot}$ star at 13 au \citep{Waisberg23c}. It was believed to have been the brightest star in the night sky about 1 Myr ago but this claim was based on an misleadingly low proper motion that was corrupted by the outer orbital motion. 

\end{enumerate}

The fourth early-type (B8III) member of the Teapot is Phi Sagittarii (HIP 92041, HD 173300, HR 7039, 27 Sgr), for which the IAU Working Group on Star Names very recently (2026/08/13) approved the proper name Nandou after the Southern Dipper mansion in the Chinese constellation system. Phi Sgr was shown to also be a binary in \cite{Waisberg23b}, where we reported the detection of a companion with a K band flux ratio of 6\% at a projected separation $\rho=17.7 \text{ mas}$ in a VLTI/GRAVITY observation. Here, we combine follow-up VLTI/GRAVITY observations, archival spectroscopic observations, past radial velocity measurements, photomety and Hipparcos astrometric data in order to solve for its orbital and stellar parameters. 

This paper is organized as follows. The observations and data are described in Section \ref{sec:observations}. Results can be found in Section \ref{sec:results}, including the spectral and photometric analysis and orbital parameters determination. Section \ref{sec:discussion} discusses the evolutionary mass of the primary as well as the future evolution of the system. A brief conclusion follows in Section \ref{sec:conclusion}. 

\section{Observations}
\label{sec:observations}

\subsection{VLTI/GRAVITY}

Phi Sgr was observed with VLTI/GRAVITY \citep{GRAVITY17} five times in 2024 over a period of about 2 months. All the observations were performed with the four 1.8-m Auxiliary Telescopes (ATs) in single-field mode, wherein half of the photons were used to fringe track at low spectral resolution (R=22) and the other half were integrated in 10-seconds exposures in high spectral resolution (R=4,000). Each observation consists of two or three files with 16 exposures each. The seeing, baseline configuration and corresponding maximum angular resolution in the working near-infrared K band can be found in Table \ref{table:vlti_observations}. In all epochs the star HD 174630 (spectral type K0III; angular diameter of 0.73 mas) was observed as interferometric calibrator. The data were reduced with the ESO GRAVITY pipeline v.1.4.0 \citep{Lapeyrere14}. Each file consists of of a set of squared visibilities for each of the six baselines and closure phases for each of the four triangles. We always make use of the science (high spectral resolution) data for the scientific analysis. 

\begin{table*}
\centering
\caption{\label{table:vlti_observations} VLTI/GRAVITY observations of Phi Sagittarii and best-fit binary model parameters.}
\begin{tabular}{cccccccccc}
\hline \hline
\shortstack{date\\MJD} & exposure time & \shortstack{seeing\\@500 nm\\(")} & \shortstack{AT\\configuration} & \shortstack{$B_{\mathrm{proj,max}}$ (m) \\$\theta_{\mathrm{max}}$ (mas)} & \shortstack{$\frac{f_{\mathrm{B}}}{f_{\mathrm{A}}}$ (\%)} & \shortstack{$\Delta \alpha_*$\\(mas)} & \shortstack{$\Delta \delta$\\(mas)} & \shortstack{$\theta_A$\\(mas)} \\ [0.3cm]

\shortstack{2017-07-19\\57953.21} & 30s x 10 x 2 & 1.5 & A0-G1-J2-J3 & \shortstack{132.3\\3.4} & $6.25\pm0.17$ & $15.80\pm0.03$ & $7.93\pm0.03$ & - \\ [0.3cm]

\shortstack{2017-07-20\\57954.08} & 30s x 10 x 2 & 0.6 & A0-G1-J2-J3 & \shortstack{132.3\\3.4} & $5.84\pm0.12$ & $15.72\pm0.03$ & $8.10\pm0.03$ & - \\ [0.3cm]

\shortstack{2024-05-18\\60448.29} & 10s x 16 x 2 & 0.8 & A0-G2-J2-J3 & \shortstack{132.0\\3.4} & $6.27\pm0.02$ & $-8.33\pm0.03$ & $-9.27\pm0.03$ & $0.65\pm0.02$ \\ [0.3cm]

\shortstack{2024-06-09\\60470.15} & 10s x 16 x 2 & 1.45 & A0-B5-J2-J6 & \shortstack{201.8\\2.2} & $6.88\pm0.36$ & $-3.68\pm0.03$ & $-8.65\pm0.03$ & $0.74\pm0.02$ \\ [0.3cm]

\shortstack{2024-06-19\\60480.39} & 10s x 16 x 3 & 0.9-1.2 & A0-B2-D0-C1 & \shortstack{28.9\\15.7} & $6.43\pm0.33$ & $-1.30\pm0.08$ & $-7.55\pm0.08$ & - \\ [0.3cm]

\shortstack{2024-07-04\\60495.28} & 10s x 16 x 2 & 0.45 & A0-G1-J2-K0 & \shortstack{110.9\\4.1} & $6.46\pm0.02$ & $2.603\pm0.03$ & $-5.607\pm0.03$ & $0.63\pm0.02$ \\ [0.3cm]

\shortstack{2024-07-26\\60517.22} & 10s x 16 x 2 & 0.7 & A0-B2-D0-C1 & \shortstack{32.3\\14.0} & $6.53\pm0.01$ & $7.86\pm0.08$ & $-2.08\pm0.08$ & - \\ [0.3cm]

\hline
\end{tabular}
\end{table*}

In addition, we also make use of the two VLTI/GRAVITY observations of Phi Sgr in 2017 reported in \cite{Waisberg23b}, which are included in Table \ref{table:vlti_observations} for convenience. 

\subsection{Spectra}
\label{obs:spectra}

We made use of five archival ``ready-to-use'' reduced spectra of Phi Sgr. Two epochs were taken in 2011 and 2014 with the HERMES spectrograph on the 1.2-meter \textit{Mercator} telescope and were downloaded from the MELCHIORS library \citep{Royer2024}. They have a wavelength coverage of 3760-9000{\AA} and an average spectral resolution $R=\frac{\lambda}{\Delta\lambda}=85,000$. Three epochs were taken in 2014 and 2015 with the ESPaDOns spectropolarimeter \citep{Donati2006} and were downloaded from the PolarBase archive \citep{Petit2014}. They have a wavelength coverage 3700-10400{\AA} and a spectral resolution $R=68,000$. For the ESPaDOns epochs in which more than one spectrum were taken in succession, we averaged the spectra weighted by their SNR. Table \ref{table:spectra} provides further details of the spectral observations including dates and exposure times. 

\begin{table}
\centering
\caption{\label{table:spectra} Details of spectral observations of Phi Sagittarii.}
\begin{tabular}{cccc}
\hline \hline
\shortstack{date\\MJD} & Instrument & Exposure time & \shortstack{$v_{\mathrm{z,A}}$\\($\text{ km}\text{ s}^{-1}$)} \\ [0.3cm]

\shortstack{2011-08-25\\55798.91} & HERMES & 100 s x 1 & $0.1\pm0.3$ \\ [0.3cm]

\shortstack{2014-06-22\\56830.11} & HERMES & 21 s x 1 & $24.7\pm0.3$ \\ [0.3cm]

\shortstack{2014-09-08\\56908.35} & ESPaDOns & 120 s x 3 & $17.6\pm0.3$ \\ [0.3cm]

\shortstack{2014-09-10\\56910.28} & ESPaDOns & 120 s x 4 & $17.4\pm0.3$ \\ [0.3cm]

\shortstack{2015-04-03\\57115.64} & ESPaDOns & 120 s x 1 & $27.0\pm0.3$ \\ [0.3cm]

\hline
\end{tabular}
\end{table}

All the reduced spectra had already been corrected to the solar system barycentric frame. The spectra were normalized by fitting a cubic spline through carefully selected continuum regions. All the wavelengths are in air. 

\subsection{Past radial velocities}

We used the Bibliography of Stellar Radial Velocities \citep{Abt1972} to find past radial velocity measurementes of Phi Sgr, which are collected in Table \ref{table:past_rv} for convenience. \cite{Lee1910} reports nine measurements between 1903 and 1910. \cite{Frost1926} reports the same measurements as in \cite{Lee1910} with one additional measurement in 1915. \cite{Campbell1928} reports six measurements between 1904 and 1909. There are no errorbars reported in these references. We estimated a typical error of $3 \text{ km}\text{ s}^{-1}$ based on the orbital fit described below. 

\subsection{Hipparcos IAD}

We downloaded the Hipparcos \citep{Hipparcos97} Intermediate Astrometric Data (IAD) for Phi Sgr using the public python package \texttt{htof} \citep{Brandt2021}. The data consists of the 1D abscissae residuals from two independent consortia, namely NDAC (Northern Data Analysis Consortium) and FAST (Fundamental Astronomy by Space Techniques). 

There is a second set of abscissae values from the Hipparcos re-reduction \citep{vanLeeuwen07}, which applied a more aggressive model for satellite attitude correction. This led to a reduction of the errors at the expense of possibly ``fitting out'' binary orbital motion \citep[see e.g.][for details]{Brandt2023}. We downloaded the corresponding abscissae from the 2014 Java tool provided by the ESA website. 

For both datasets we reconstructed the original abscissae values $\nu$ based on the abscissa residuals res, the reported parallax $p$ and the reported proper motions $(\mu_{\alpha_*},\mu_{\delta})$ 

\begin{align}
\nu = \mathrm{res} + \mu_{\alpha_*} (t-t_0) \cos \psi + \mu_{\delta} (t-t_0) \sin \psi + p * \mathrm{parf}
\end{align}

\noindent where $\psi$ is the scan angle, parf is the parallax factor and $t_0=1991.25$ is the reference epoch. 

We note that the IAD file from the Hipparcos re-reduction provides $t$, $\cos \psi$, $\sin \psi$ and parf. 

The original Hipparcos reduction directly provides 

\begin{align}
\cos \psi = \frac{d \nu}{d \alpha_*} \\
\sin \psi = \frac{d \nu}{d \delta} \\ 
\mathrm{parf} = \frac{d \nu}{d p}
\end{align}

\noindent while the epoch can be calculated 

\begin{align}
t = \frac{d\nu}{d\alpha_*}\frac{d\nu}{d \mu_{\alpha_*}} + \frac{d\nu}{d \delta}\frac{d \nu}{d \mu_{\delta}}
\end{align}

\section{Results}
\label{sec:results}

\subsection{Separation vectors from interferometry}

The interferometric data for each epoch were fit with a binary model \citep[detailed in][]{Waisberg23} consisting of the K band flux ratio $\frac{f_{B}}{f_{A}}$ and the projected separation of the secondary B relative to the primary A in the East and North directions $(\Delta \alpha_*, \Delta \delta)_{\mathrm{B,A}}$. The angular diameter of the secondary was fixed to $\theta_B = 0.2 \text{ mas}$ based on its estimated radius; it cannot be meaningfully constrained by the interferometric data since it is well below the maximum resolution. Meanwhile, the angular diameter of the primary $\theta_A$ is large enough to be measured in some of the epochs with the highest spatial resolution. 

\begin{figure*}[]
\includegraphics[width=2.2\columnwidth]{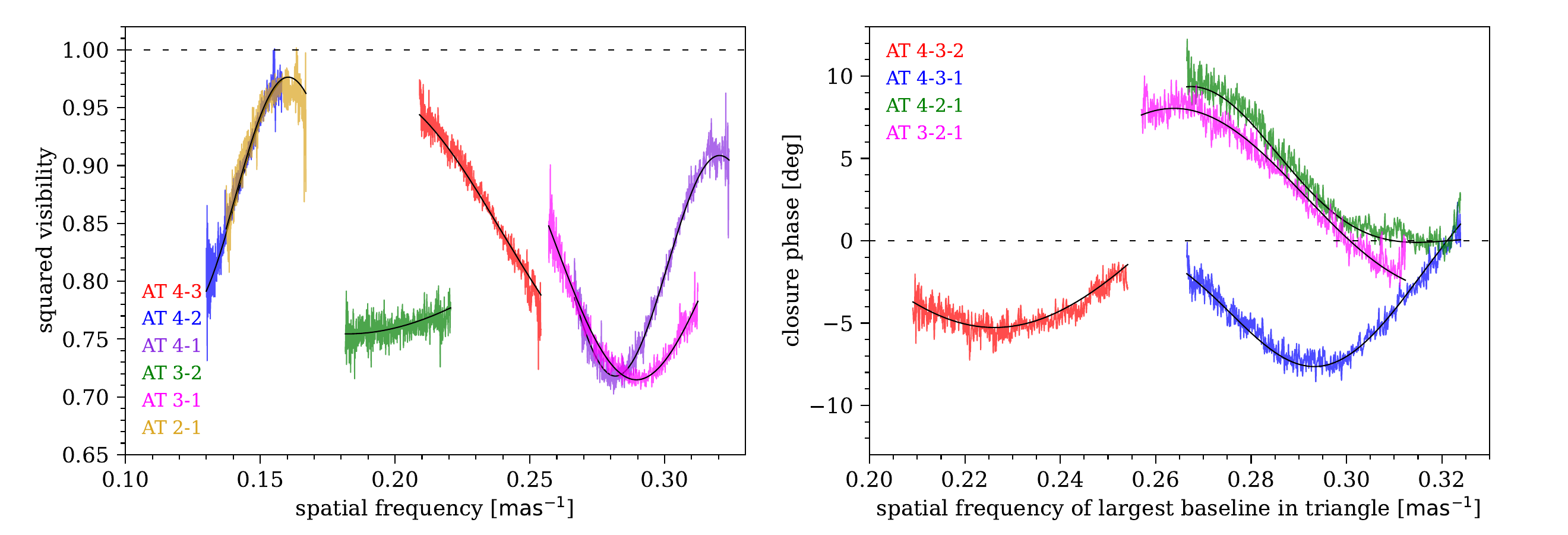}
\caption{VLTI/GRAVITY data for Phi Sagittarii (colored) and best fit binary model (solid black) for epoch 2024-05-18. The expected values for a single unresolved star are shown in dashed lines.}
\label{fig:gravity_data}
\end{figure*}

Figure \ref{fig:gravity_data} shows an example of the interferometric data (colored) and best fit binary model (solid black) for epoch 2024-05-18. 

The best fit parameters for each epoch are reported in Table \ref{table:vlti_observations}. The errors were estimated from the scatter between the results of different files over the same night. We found a typical astrometric error of 30 $\mu$as for the data with high spatial resolution and 80 $\mu$as for the data with lower spatial resolution. The weighted mean and error for the K band flux ratio is $f_{\mathrm{K}}=6.5\pm0.1\%$. The angular diameter of the primary is $\theta_A = 0.67 \pm 0.05 \text{ mas}$. 

\subsection{Spectral analysis and radial velocities}

\begin{figure*}[]
\includegraphics[width=2\columnwidth]{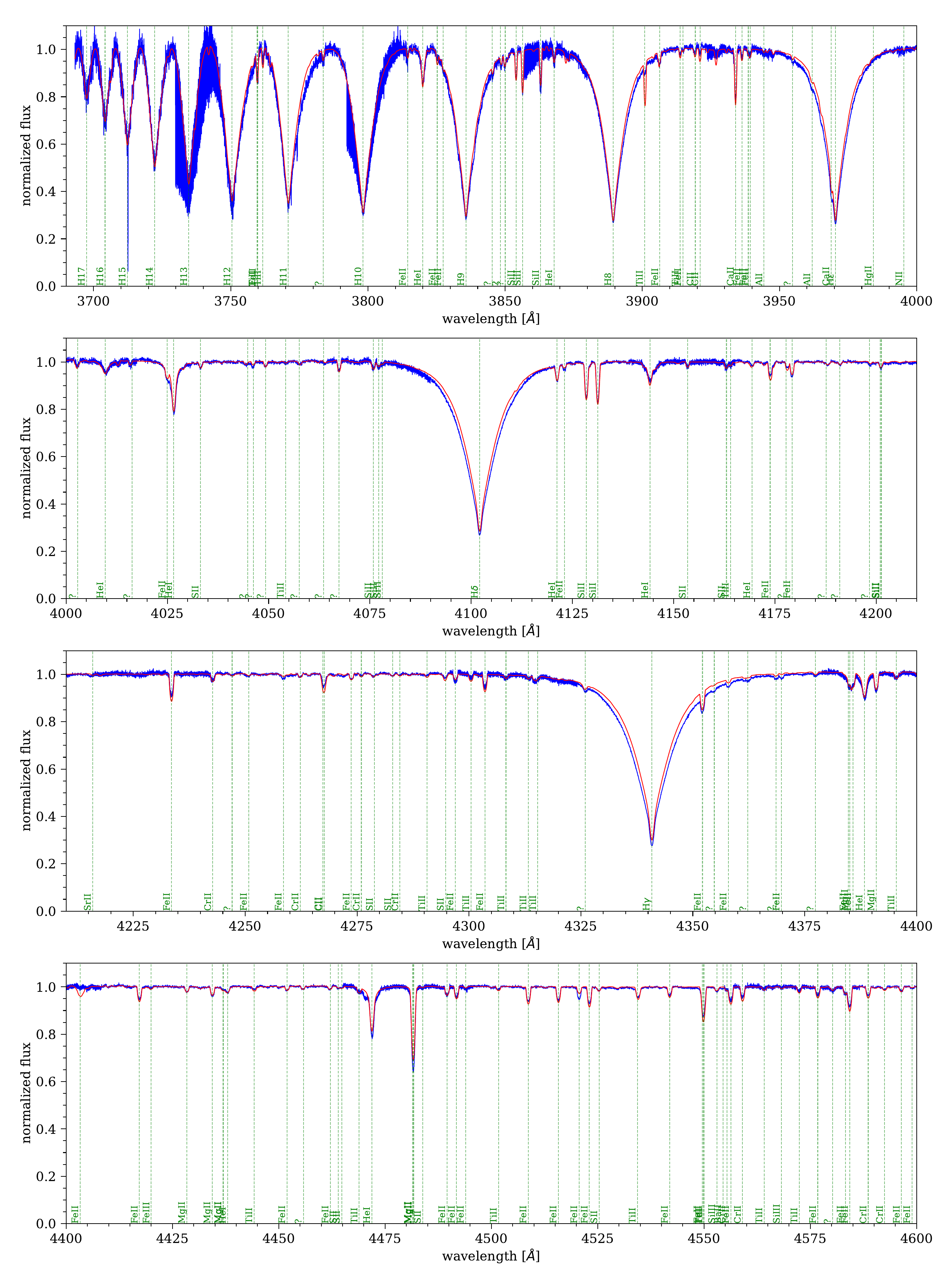}
\caption{ESPaDOns normalized spectrum of Phi Sagittarii (blue) for epoch 2015-04-03 and the best fit normalized BOSZ spectrum (red).}
\label{fig:spectrum_1}
\end{figure*}

The spectrum is totally dominated by the primary star and we did not find any evidence for spectral lines of the secondary. Figures \ref{fig:spectrum_1}, \ref{fig:spectrum_2}, \ref{fig:spectrum_3} and \ref{fig:spectrum_4} plot the normalized Espadons spectrum for epoch 2015-04-03. The identified spectral lines are marked and also listed in Table \ref{table:line_list}. 

We measured the projected rotational velocity of the primary $(v \sin i)_A = 39\pm1 \text{ km}\text{ s}^{-1}$ by finding the first zero of the Fourier transform across clean metal lines such as Mg II 4481.2{\AA}, Si II 6347.1{\AA} and Si II 6371.4{\AA}. We then compared the spectrum against instrumentally and rotationally broadened synthetic spectra from the BOSZ spectral library \citep{Bohlin2017,Meszaros2024}. The spectra are based on ATLAS9 model atmospheres \citep{Meszaros2012} and cover $T_{\mathrm{eff}}$ from 7500K to 16000K and $\log g$ from 2.0 to 5.0. We used the spectra with solar metallicity and a microturbulent velocity of $2 \text{ km}\text{ s}^{-1}$. The temperature was constrained by the ionization balance between the dominant Si II lines and the very weak Si III lines, whereas $\log g$ was constrained by the wings of the Balmer lines. We find $T_A = 12000\pm250 \text{ K}$ and $(\log g)_A \simeq 3.5$. The agreement between the real and the model spectra is very good except for a few lines (such as N I 8703{\AA}, 8712{\AA}, 8719{\AA} and O I 7771-5{\AA}, 8446{\AA}) that are strongly affected by NLTE effects (the models assume LTE). 

The radial velocities of the primary were measured by cross-correlation of the spectra with the best atmosphere model across nine carefully selected spectral regions containing strong metal lines. The error is estimated from the scatter across the different regions. The resulting radial velocities are listed in Table \ref{table:spectra}.

\subsection{Orbital parameters}
\label{subsec:orbital_fit}

In order to find the orbital parameters of $\phi$ Sgr, we start with the radial velocity data and construct a 3d grid in orbital period (150 days $< P <$ 400 days in steps of 0.1 day), eccentricity ($0 < e < 0.95$ in steps of 0.05) and time of periastron $T_p$ (in steps of 1 day). For each point in the grid, we find the optimal solution through a fast linear least squares fit on the remaining parameters, namely the radial velocity semi-amplitude of the primary $K_A$, the argument of periastron of the primary $\omega_A = \omega_B + \pi$ and the systemic velocity $\gamma$\footnote{We also added two extra parameters to account for possible systemic velocity differences between the new and the two sets of old radial velocity data. Such differences could arise from systematic errors in the old data or due to an undetected third body on a much wider orbit. We found $\gamma_1 = 13.1 \pm 1.1 \text{ km}\text{ s}^{-1}$ for the \cite{Lee1910,Frost1926} data and $\gamma_2 = 13.1 \pm 1.2 \text{ km}\text{ s}^{-1}$ for the \cite{Campbell1928} data, which are consistent with the modern measurements ($\gamma = 12.66 \pm 0.15 \text{ km}\text{ s}^{-1}$). Therefore, there is no evidence for a third body in the radial velocity data.}. We found a clear minimum near $P \simeq 240 \text{ days}$. 

We then incorporate the astrometric data by constructing another 3d grid around the optimal solution also in $P$, $e$ and $T_p$. For each point in the grid, we find the corresponding optimal Thiele-Innes elements $A,B,F,G$ (which map to the remaining four Keplerian elements, namely semi-major axis $a$, orbital inclination $i$, argument of pericenter $\omega$ and longitude of the ascending node $\Omega$) through a fast linear least squares fit to the astrometric data. We then fit (again through linear least squares) for $K_A$ and $\gamma$ using the radial velocities. This then yields an optimal solution for all the orbital parameters. Note that because we have both astrometry and radial velocities, we can break the $(\omega,\Omega) \leftrightarrow (\omega + \pi, \Omega +\pi)$ degeneracy that plagues pure astrometric solutions. For further details on the method, see for e.g. \cite{Waisberg26}. 

\begin{figure*}[]
\centering
\includegraphics[width=0.8\textwidth]{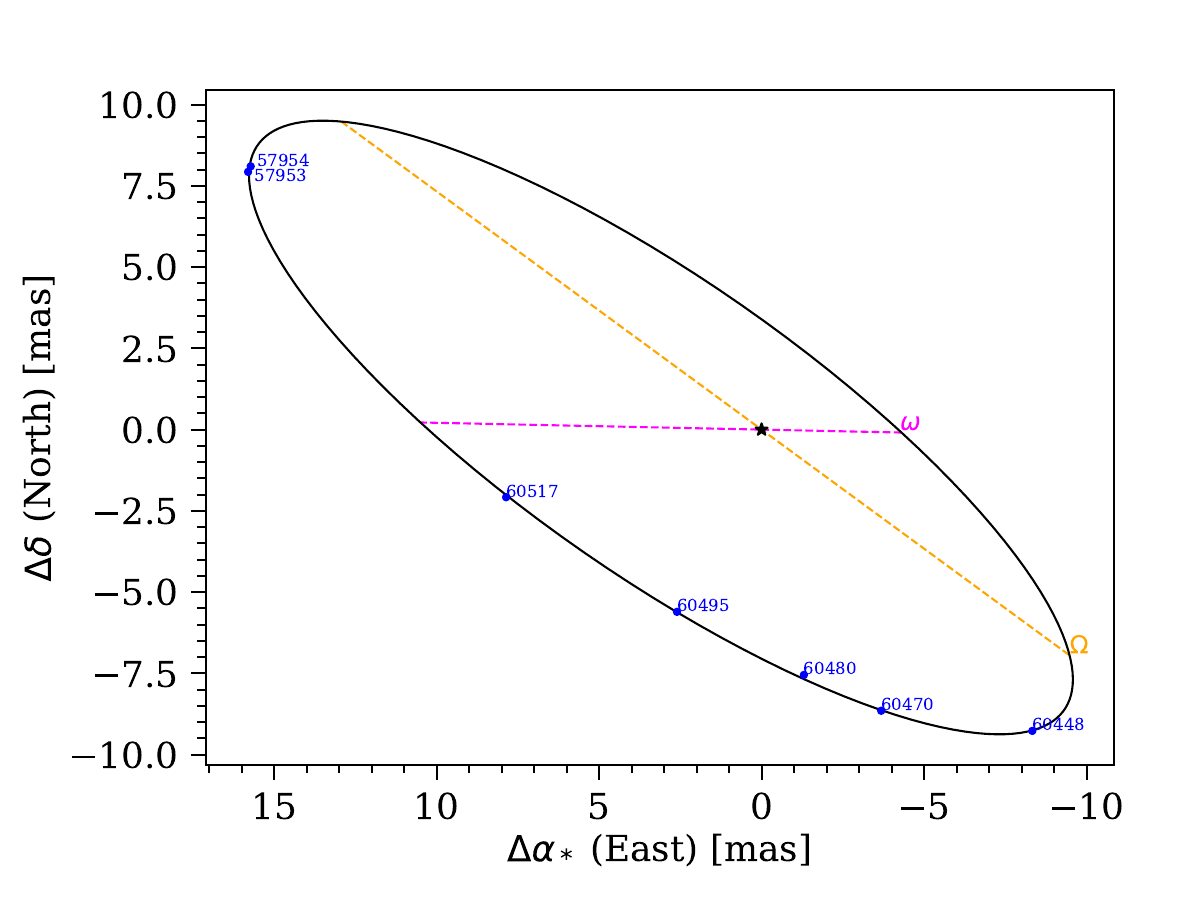} \\
\includegraphics[width=0.8\textwidth]{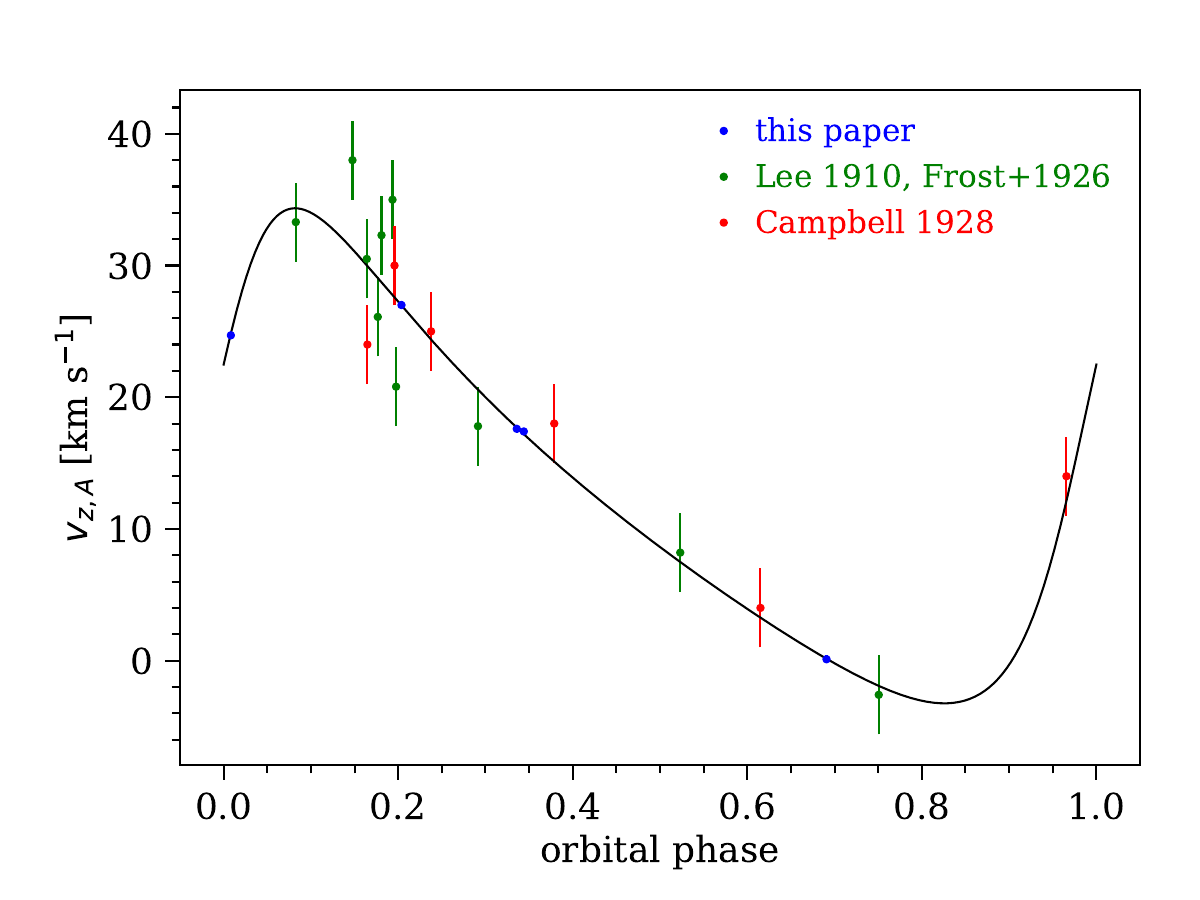}
\caption{\label{fig:orbital_fit} \textbf{Top:} VLTI/GRAVITY interferometric positions for Phi Sagittarii (blue; labeled by MJD) and best fit orbital solution (black line). The star at (0,0) marks the position of the primary. The dashed magenta and orange lines show the line of apsides and the line of nodes respectively. \textbf{Bottom:} Radial velocity data of the primary star of Phi Sagittarii (colored) and the corresponding radial velocity curve for the best fit orbital solution (black line). The data span 112 years (1903-2015) or 171 orbits.}
\end{figure*}

In a last step, we refine the orbital parameters through a non-linear least squares fit (with all parameters allowed to vary) starting from the optimal grid solution. The resulting $\chi^2=38$ for 24 degrees of freedom suggests that the average errors may have been underestimated by at most about $20\%$. Table \ref{table:orbital_fit} lists the best fit orbital parameters. The parameter errors were estimated by re-sampling the data according to the errors 5000 times and finding the 16th and 84th percentile threshold of the resulting best fit parameter distributions ($1\sigma$). Figure \ref{fig:orbital_fit} (top) shows the astrometric data (blue) and the projected best fit orbit (black) as well as (bottom) the radial velocity data for the primary star (blue) and best fit radial velocity curve (black). The full distribution plots for the orbital parameters are plotted in Figure \ref{fig:orbital_fit_corner}. There are no strong degeneracies between the parameters.

\begin{table}
\centering
\caption{\label{table:orbital_fit} Orbital and stellar parameters for Phi Sagittarii.}
\begin{tabular}{ccc}
\hline \hline

& & source \\ 

\shortstack{$T_A$\\($K$)} & $12000\pm250$ & spectrum, SED \\ [0.3cm]

\shortstack{$T_B$\\($K$)} & 7000 & see text \\ [0.3cm]

$(\log g)_A$ & 3.5 & spectrum \\ [0.3cm]

\shortstack{$(v \sin i)_A$\\($\text{ km}\text{ s}^{-1}$)} & $39\pm1$ & spectrum \\ [0.3cm]

\shortstack{$R_A$\\($R_{\odot}$)} & $5.1\pm0.4$ & SED \\ [0.3cm]

\shortstack{$R_B$\\($R_{\odot}$)} & $1.67\pm0.15$ & see text \\ [0.3cm]

\hline \\ [0.1cm]

\shortstack{$a$\\(mas)} & \shortstack{$16.44\pm0.04$} & interferometry \\ [0.3cm]

$e$ & $0.418\pm0.004$ & interferometry + RV \\ [0.3cm]

\shortstack{$i$\\(deg)} & $106.1\pm0.1$ & interferometry \\ [0.3cm]

\shortstack{$\Omega$\\(deg)} & $53.8\pm0.1$ & interferometry \\ [0.3cm]

\shortstack{$\omega_B$\\(deg)} & $111.6\pm0.2$ & interferometry + RV \\ [0.3cm]

\shortstack{$P$\\(days)} & $238.844\pm0.011$ & interferometry + RV \\ [0.3cm]

\shortstack{$T_p$\\(JD)} & $58022.3\pm0.3$ & interferometry + RV \\ [0.3cm]

\shortstack{$K_A$\\($\text{ km}\text{ s}^{-1}$)} & $18.8\pm0.3$ & RV \\ [0.3cm]

\shortstack{$\gamma$\\($\text{ km}\text{ s}^{-1}$)} & $12.66\pm0.15$ & RV \\ [0.3cm]

\hline \\ [0.1cm]

\shortstack{plx(mas)\\d(pc)} & \shortstack{$12.5\pm1.0$\\$80\pm7$} & Hipparcos IAD \\ [0.3cm]

\hline \\ [0.1cm]

\shortstack{$a$\\(au)} & $1.32\pm0.10$ & derived from above \\ [0.3cm]

$q=\frac{M_{B}}{M_{A}}$ & $0.42\pm0.01$ & derived from above \\ [0.3cm]

\shortstack{$M_{A}$\\($M_{\odot}$)} & $3.8\pm0.9$ & derived from above \\ [0.3cm]

\shortstack{$M_{B}$\\($M_{\odot}$)} & $1.6\pm0.4$ & derived from above \\ [0.3cm]

\hline
\end{tabular}
\tablenotetext{0}{Notes:}
\tablenotetext{0}{The uncertainties correspond to the 16\% and 84\% percentiles of the parameter distributions.}
\end{table}

\subsection{Parallax and dynamical masses}
\label{subsec:hipparcos}

The last obstacle to obtaining model-independent dynamical masses for $\phi$ Sgr is an accurate parallax. The original Hipparcos reduction \citep{Hipparcos97} reported a parallax $p=14.14\pm0.88 \text{ mas}$. The new Hipparcos reduction \citep{vanLeeuwen07} reported $p=13.63\pm0.19 \text{ mas}$. Gaia DR2 \citep{Gaia18} reported $p=12.16\pm0.59 \text{ mas}$ while Gaia DR3 \citep{Gaia21} did not report a parallax for $\phi$ Sgr. 

The scatter in the parallax values is likely caused by the unmodeled orbital motion of the photocenter with the period that is a significant fraction of one year. The photocenter traces the same orbit as the primary star but with a slightly reduced semi-major axis:

\begin{align}
a_{\mathrm{phot}} = a \left ( \frac{q}{1+q} - \frac{f}{1+f} \right )
\end{align}

\noindent where $q = \frac{M_B}{M_A}$ is the mass ratio and $f=\frac{f_B}{f_A}$ is the flux ratio in the Hipparcos band. For a given parallax, the orbital solution in \ref{subsec:orbital_fit} gives the dynamical stellar masses and, assuming the secondary is a main-sequence star, the flux ratio in the visual band (similar to the Hipparcos band) can be estimated based on standard calibrated mass-magnitude relations and the total visual magnitude of $\phi$ Sgr ($V_T=3.13$). Table \ref{table:photocenter} reports the corresponding values for parallaxes of 12, 13 and 14 mas. The resulting $a_{\mathrm{phot}}$ are 4.3, 4.8 and 5.3 mas, so it is quite plausible that the unmodeled orbital motion of the photocenter indeed affects the reported parallaxes. 

\begin{table}
\centering
\caption{\label{table:photocenter} Masses, visual band flux ratio and semi-major axis of the photocenter as a function of parallax.}
\begin{tabular}{ccccc}
\hline \hline
\shortstack{parallax\\(mas)} & $q=\frac{M_B}{M_A}$ & \shortstack{$M_B$\\($M_{\odot}$)} & $f_V = \frac{f_{V,B}}{f_{V,A}}$ & \shortstack{$a_{\mathrm{phot}}$\\(mas)} \\ [0.3cm]

12 & 0.40 & 1.72 & 0.028 & 4.3 \\ [0.3cm]

13 & 0.45 & 1.47 & 0.020 & 4.8 \\ [0.3cm]

14 & 0.50 & 1.26 & 0.012 & 5.3 \\ [0.3cm]

\hline
\end{tabular}
\end{table}

We attempted to obtain an unbiased value for the parallax by fitting the reconstructed abscissae from the Hipparcos IAD with a model that includes the photocenter orbital motion in addition to the canonical 5-parameter model (reference positions, proper motions and parallax). The orbital parameters (except the semi-major axis) were fixed to the best fit orbital solution. Note that the argument of the pericenter of the photocenter $\omega_{\mathrm{phot}} = \omega_A = \omega_B + \pi$. We also fixed the proper motion to that calculated from the Gaia DR3-Hipparcos positional difference over 2016.0-1991.25=24.75 years, namely $(v_{\alpha_*},v_{\delta}) = (52.1, -0.44) \text{ mas}\text{ yr}^{-1}$. Therefore, the free parameters were the parallax $p$, the photocenter semi-major axis $a_{\mathrm{phot}}$ and a 2d offset $(\Delta \alpha_*, \Delta \delta)$ from the reference position. 

\begin{figure}[]
\centering
\includegraphics[width=0.5\textwidth]{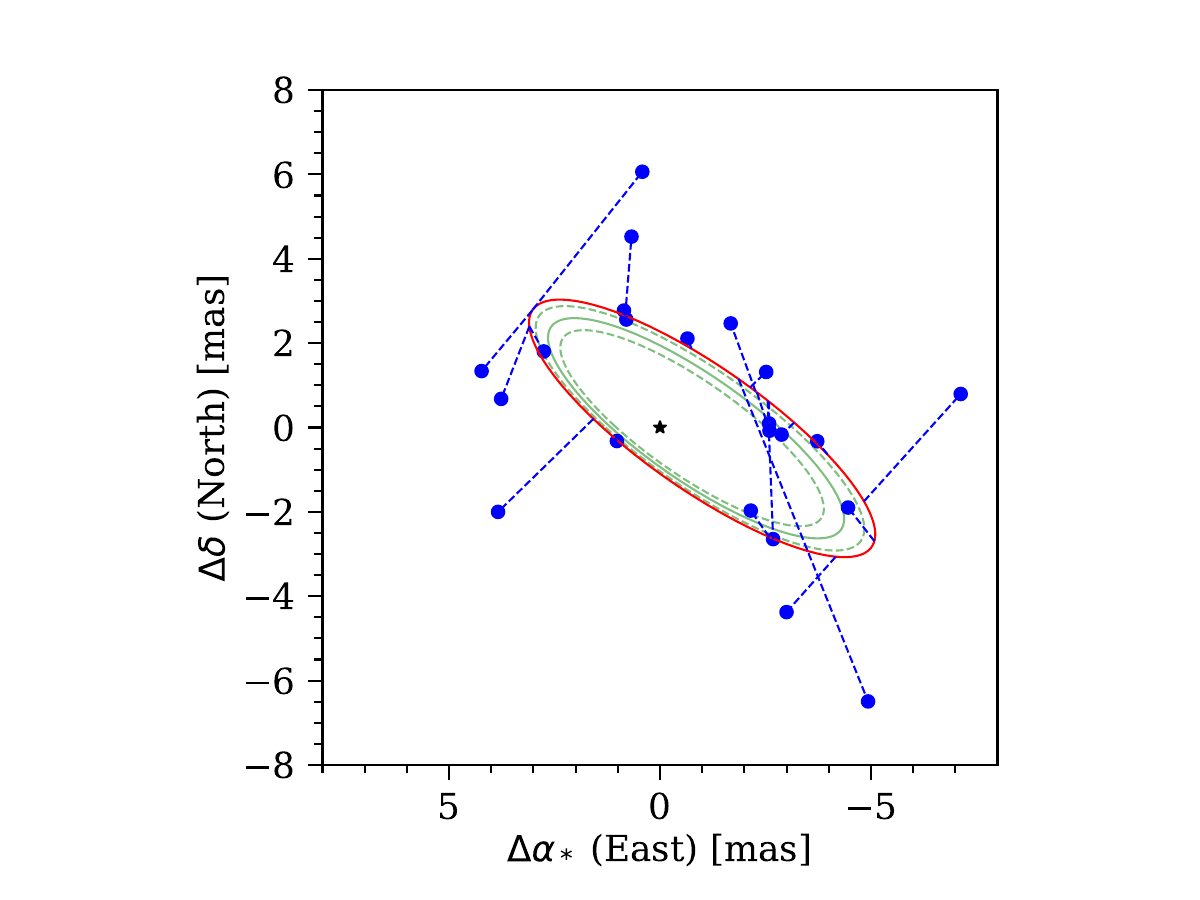}\\
\includegraphics[width=0.5\textwidth]{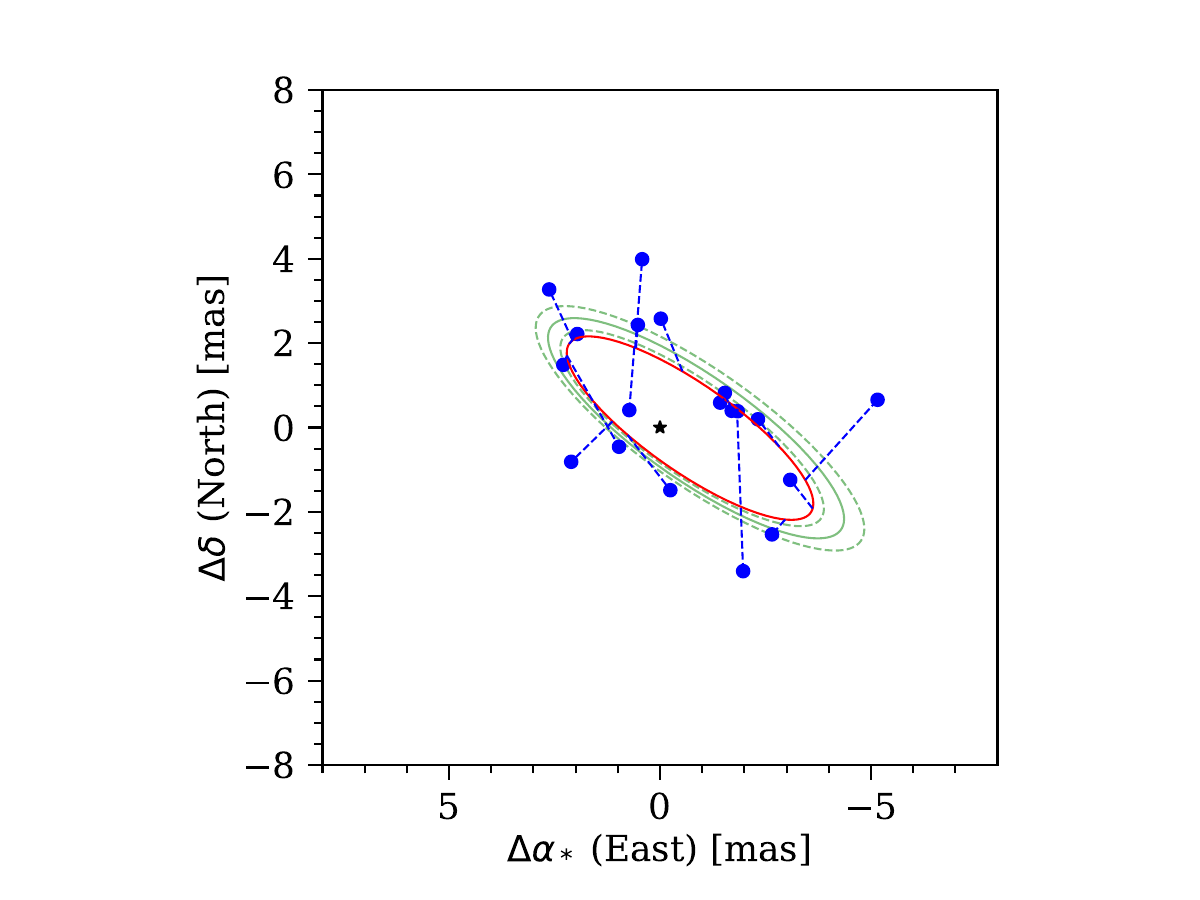}\\
\includegraphics[width=0.5\textwidth]{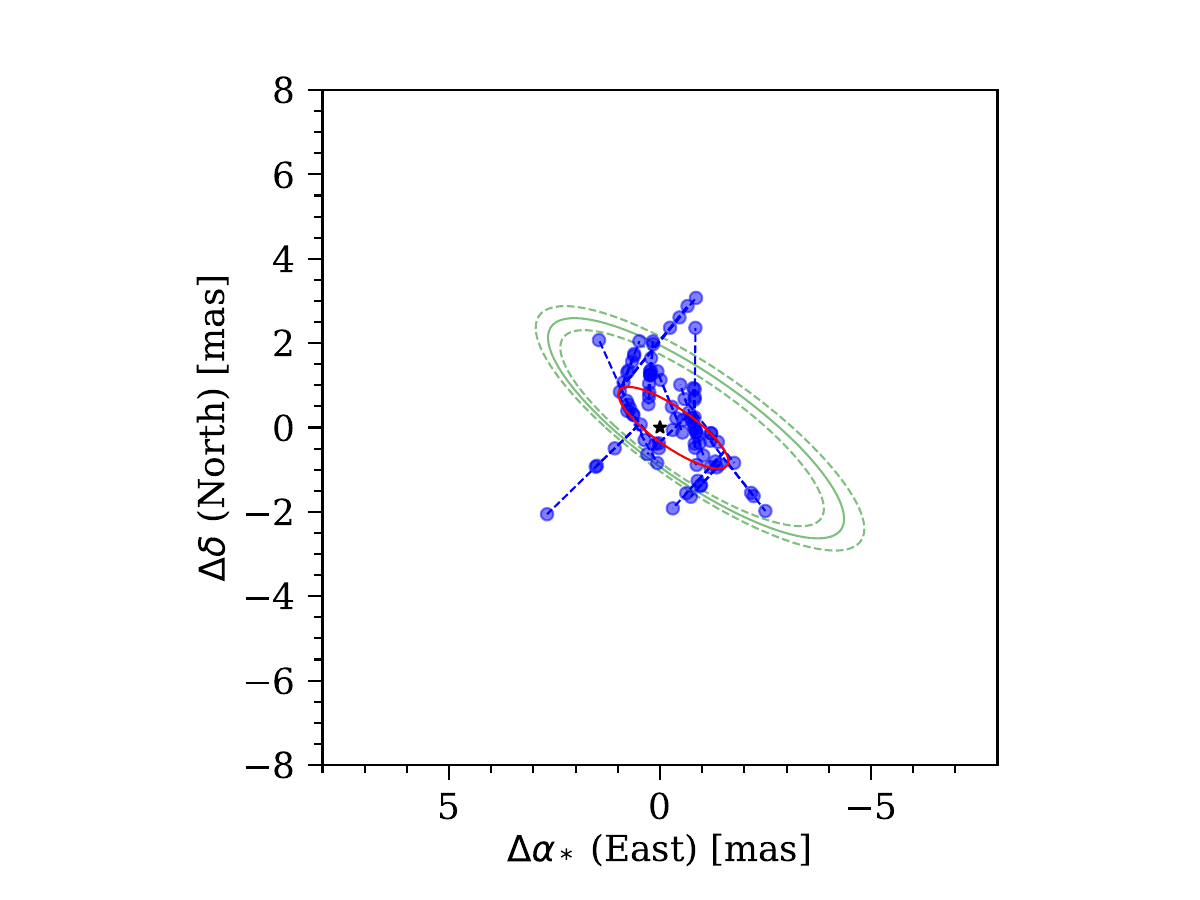}
\caption{\label{fig:hipparcos_fits} Abscissae residuals (blue) of the Hipparcos Intermediate Astrometric Data (IAD) against the best fit photocenter orbit (red). The expected photocenter orbits for parallaxes $\mathrm{plx}=12.5\pm1.0 \text{ mas}$ ($a_{\mathrm{phot}} = 4.55\pm0.50 \text{ mas}$) are shown in green. \textbf{Top}: original Hipparcos 1997 data from the NDAC consortium. \textbf{Middle}: original Hipparcos 1997 data from the FAST consortium. \textbf{Bottom}: Hipparcos re-reduction 2007 data.}
\end{figure}

Table \ref{table:hipparcos_fits} reports the results. Figure \ref{fig:hipparcos_fits} shows the corresponding orbits in a way that is typically used to illustrate the Hipparcos one dimensional abscissae \citep[e.g.][]{Torres2007,Sahlmann2011}. The red line shows the sky-projected photocenter orbit, where $a_{\mathrm{phot}}$ is set to the best fit value. The blue points show the abscissae values in the following way: each abscissae is connected to the corresponding point on the orbit at its given epoch through the dashed lines and the distance between them is given by the difference between the model and the measured abscissae projected along the scan angle direction. The green lines show the expected photocenter orbit for parallaxes $\mathrm{plx}=12.5\pm1.0 \text{ mas}$ ($a_{\mathrm{phot}} = 4.55\pm0.50 \text{ mas}$). 

\begin{table*}
\centering
\caption{\label{table:hipparcos_fits} Results for Hipparcos Intermediate Astrometric Data fits.}
\begin{tabular}{cccccc}
\hline \hline

dataset & \shortstack{parallax\\(mas)} & \shortstack{$a_{\mathrm{phot}}$\\(mas)} & \shortstack{$\Delta \alpha_*$\\(mas)} & \shortstack{$\Delta \delta$\\(mas)} & $\chi^2$ / dof \\ [0.3cm]

\shortstack{Hip 1997\\NDAC} & $13.0\pm1.1$ & $5.3\pm1.3$ & $2.1\pm1.1$ & $-0.62\pm0.55$ & 18.6 / 18 \\ [0.3cm]

\shortstack{Hip 1997\\FAST} & $12.1\pm0.9$ & $3.8\pm1.0$ & $1.6\pm0.9$ & $-0.48\pm0.47$ & 10.3  / 15 \\ [0.3cm]

Hip 2007 & $12.98 \pm 0.25$ & $1.7\pm0.3$ & $0.86 \pm 0.23$ & $0.02\pm0.14$ & 159 / 84 \\ [0.3cm]

\hline
\end{tabular}
\end{table*}

The photocenter semi-major axes from the original Hipparcos IAD -- $a_{\mathrm{phot}} = 5.3 \pm 1.0 \text{ mas}$ for NDAC and $a_{\mathrm{phot}} = 3.8 \pm 1.0 \text{ mas}$ for FAST -- are consistent with the range of expected $a_{\mathrm{phot}}$ albeit the errors are relatively large. We adopt a conservative parallax $p = 12.5 \pm 1.0 \text{ mas}$ based on the original Hipparcos IAD. 

Meanwhile, $a_{\mathrm{phot}} = 1.7 \pm 0.3 \text{ mas}$ for the Hipparcos re-reduction IAD is much smaller than the expected range. For the Hipparcos photocenter to be so small, the V band flux ratio would have to be much larger (26\%) than the expected 2\%. Given $V_T=3.14$ and $K=3.28$ for Phi Sgr, this would imply $V=4.6$ and $K=6.3$ for the companion (the latter comes from the interferometric flux ratio). Not only do we not see any evidence for spectral lines of the secondary in the optical spectra but it is physically impossible for any thermal source to be so blue ($V-K=-1.7$). Therefore, we conclude that the orbital motion was significantly erased by the more aggressive processing of the Hipparcos re-reduction data. This issue has been previously documented for e.g. in \cite{Brandt2021,Brandt2023}. Because of it, we refrain from using the formally more precise parallax from the Hipparcos 2007 IAD. 

The error is our dynamical masses $M_A=3.8 \pm 0.9 M_{\odot}$ and $M_B=1.6 \pm 0.4 M_{\odot}$ are ultimately limited by the parallax error. 

The 2MASS K band magnitude of Phi Sgr is $K=3.28\pm0.32$ \citep{Skrutskie2006}. For a distance of 80 pc and using the interferometric K band flux ratio, the absolute K band magnitudes are $M_{K,A} = -1.2 \pm 0.3$ and $M_{K,B} = 1.8 \pm 0.3$. The expected absolute K band magnitude for a main sequence star of mass $M_B = 1.6 M_{\odot}$ is $M_K\simeq1.8$. Therefore, the companion is consistent with being a main-sequence star. 

\subsection{Photometric radius and temperature}
\label{subsec:radius}

We measured the temperature and radius of the primary star using space-based photometry of $\phi$ Sgr, namely the four ultraviolet bands from TD-1 \citep{Thompson1978}, the $V_T$ and $B_T$ bands from Tycho2 \citep{Hog2000}, near-infrared $J$, $H$ and $K$ bands from 2MASS \citep{Skrutskie2006} and W1, W2, W3 and W4 bands from WISE \citep{Wright2010}. We refrained from using Gaia photometry for such a bright star (for e.g. its DR2 and DR3 photometry are discrepant by 0.1-0.2 mag). 

We used the BOSZ templates with $\log g=3.5$ and for each find the best fit primary radius. The interstellar extinction towards $\phi$ Sgr is negligible. We note that the contribution of the secondary to the SED is negligible within the photometric errorbars. In any case, we included it with a BOSZ template appropriate for a $1.6 M_{\odot}$ MS star ($T_{\mathrm{eff,B}}=7250 \text{ K}$, $\log g=4.0$) and a radius fixed by the K band interferometric flux ratio (where both stars are in the Rayleigh-Jeans limit):

\begin{align}
\frac{R_B}{R_A} \simeq \left ( \frac{f_{K,B}}{f_{K,A}} \frac{T_A}{T_B} \right )^{1/2}
\end{align}

The model fluxes over each filter were calculated using the public module \texttt{pyphot} \citep{Fouesneau20215}. Filter profiles for the TD-1 bands (not included in \texttt{pyphot}) were downloaded from the SVO Filter Profile Service \citep{Rodrigo2024}. 

\begin{figure}[]
\centering
\includegraphics[width=0.5\textwidth]{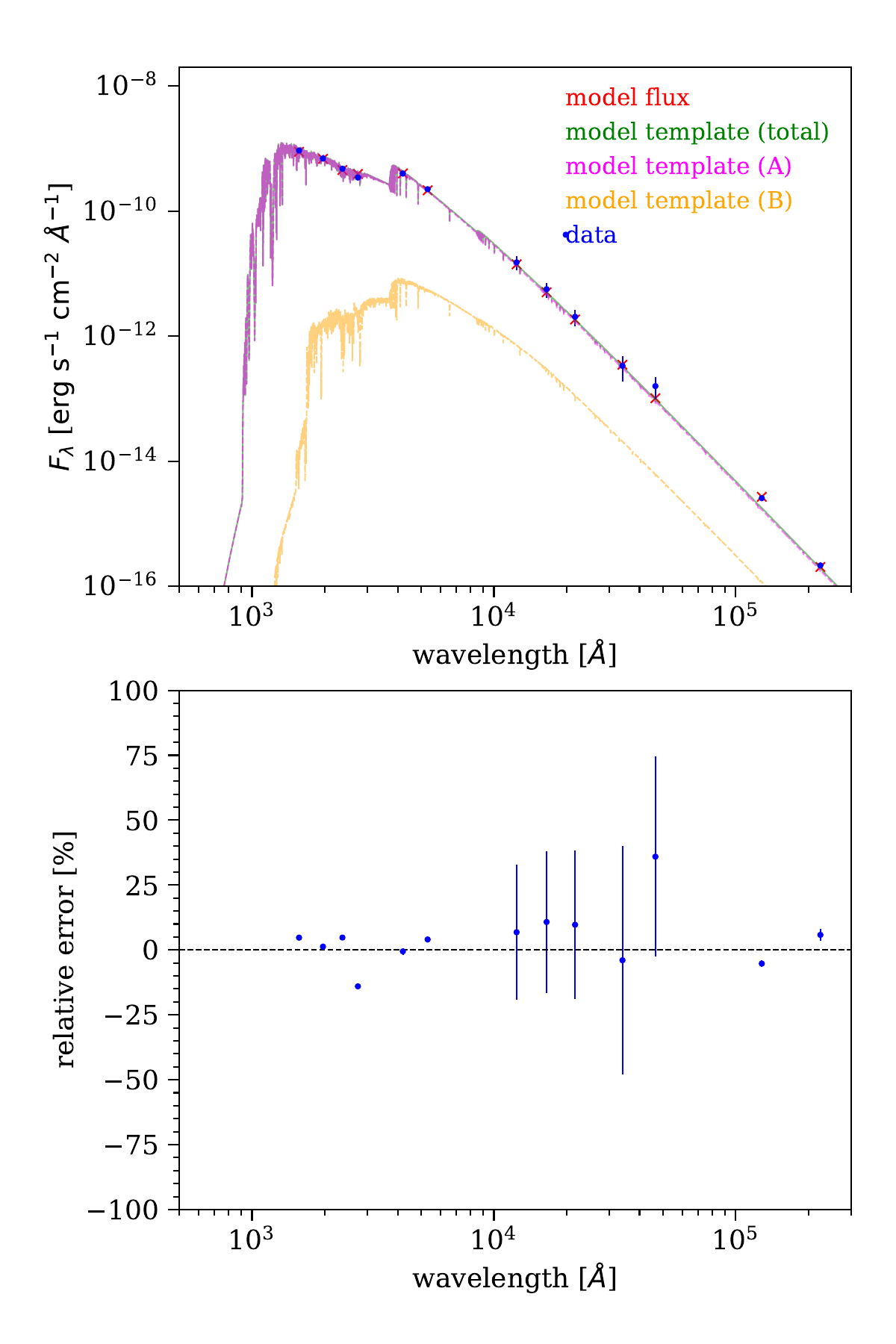}
\caption{\label{fig:photometry_fit} \textbf{Top:} Photometric data of Phi Sgr from TD-1, Tycho, 2MASS and WISE together with the best fit model template from BOSZ. \textbf{Bottom:} Corresponding residuals.}
\end{figure}

The best fit template had $T_{\mathrm{eff,A}}= 12 \text{ kK}$ (i.e. the photometric temperature agrees with the spectroscopic temperature) and the corresponding radii are $R_A = 5.1 \pm 0.4 R_{\odot}$ and $R_B \simeq 1.67 \pm 0.15 R_{\odot}$. The radii errorbars are dominated by the parallax uncertainty. The corresponding angular diameter of the primary $\theta_{A} = 0.59 \text{ mas}$ is rouughly consistent with the one directly measured from interferometry. The photometric data, model template and residuals are plotted in Figure \ref{fig:photometry_fit}. 

\section{Discussion}
\label{sec:discussion}

\subsection{Evolutionary mass}

\begin{figure}[H]
\centering
\includegraphics[width=\columnwidth]{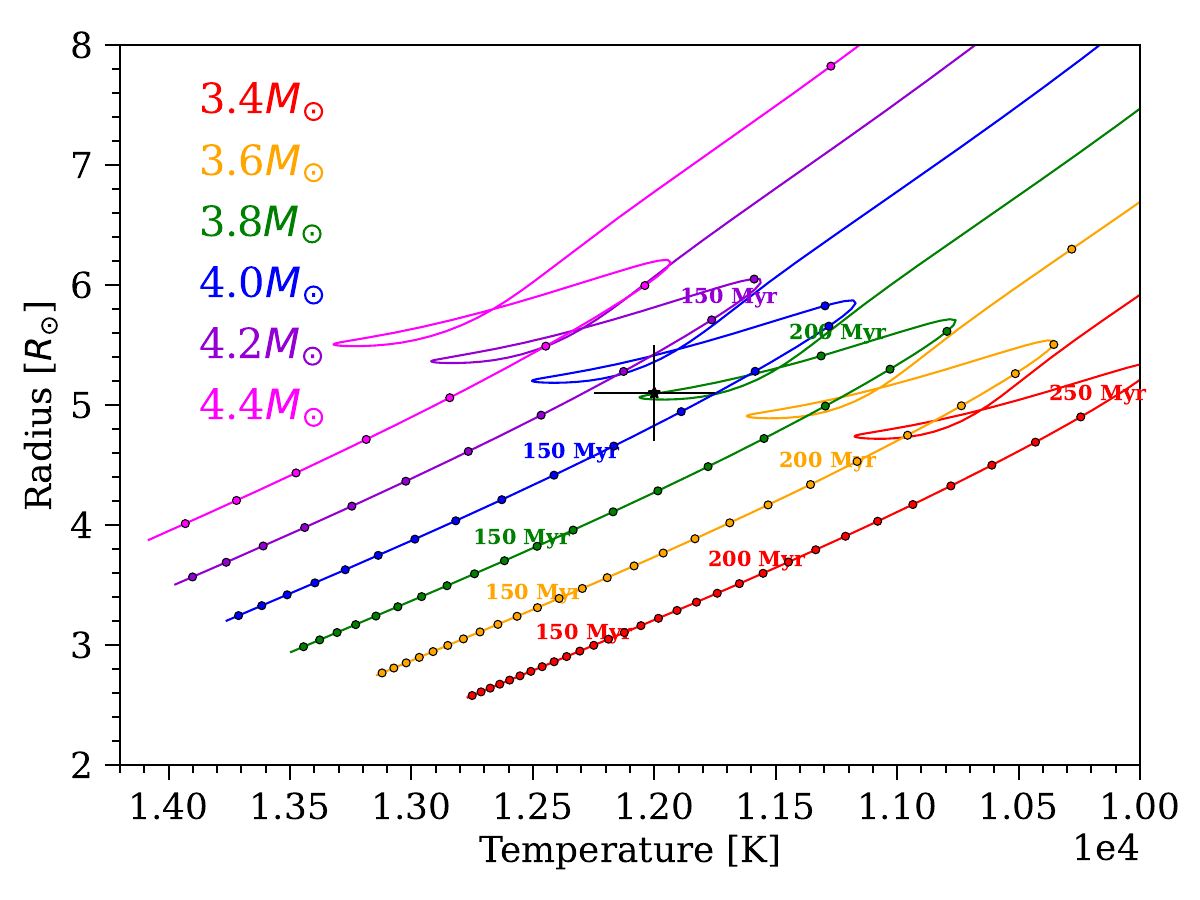}
\caption{\label{fig:evolutionary_tracks} \texttt{MIST} stellar evolutionary tracks for a range of different masses and initial rotation of 0.4 times critical. The dots mark 5 Myr intervals. The measured parameters for Phi Sagittarii A are marked in black.}
\end{figure}

In Figure \ref{fig:evolutionary_tracks} we plot evolutionary models from \texttt{MIST} \citep{Dotter16,Choi16,Paxton11,Paxton13,Paxton15} for stellar masses $3.4-4.4 M_{\odot}$ and age $\tau > 100 \text{ Myr}$. The dots mark 5 Myr intervals. The models have solar metallicity and a initial rotational velocity of 0.4 times the critical velocity. The temperature and radius of Phi Sgr A are plotted in black and imply that it is most likely a $\tau \sim 150 \text{ Myr}$ $4.0-4.2 M_{\odot}$ star (the mass ratio then implies $M_B \simeq 1.68-1.76 M_{\odot}$) near the end of the main-sequence phase. In principle it could also be a $\tau \sim 200 \text{ Myr}$, $3.8 M_{\odot}$ star in the main-sequence hook just after finishing core hydrogen burning but since this is an extremely brief phase it is very unlikely that we would observe it. Because the precision of the dynamical mass is currently limited by the relatively uncertain parallax, Phi Sgr cannot yet be used to test the accuracy of the evolutionary mass. 

\subsection{Future evolution}

The Roche lobe radius of the primary star for a circular and synchronized binary with a mass ratio $q=\frac{M_B}{M_A}=0.42$ is $\frac{R_{\mathrm{L,A}}}{a} = 0.45$, where $a$ is the semi-major axis \citep{Eggleton1983}. The primary in $\phi$ Sgr would fill its pseudo-Roche lobe at periastron once it has a radius $R \sim 0.45 a (1-e) \simeq 70 R_{\odot}$. According to the \texttt{MIST} models a $M=3.8-4.2 M_{\odot}$ star reaches a maximum radius $R\simeq 60-70 R_{\odot}$ in the red giant (RG) phase so would at best barely fill its pseudo-Roche lobe. It is therefore safe to say that the primary will be deep in the RG phase before any significant mass transfer could happen. By then it already has a deep convective envelope that is very efficient in dissipating the orbital energy through equilibrium tides \citep[e.g.][]{Zahn1977}. In other words, the binary should circularize before mass transfer starts and avoid eccentric Roche lobe overflow. By conservation of angular momentum\footnote{Note that the spin angular momenta are negligible compared to the orbital angular momentum even if the primary were rotating near critical velocity.} the circularized semi-major axis will be 

\begin{align}
a_f = a_i (1-e_i^2) \simeq 1.05 \text{ au} 
\end{align}

\noindent so that the primary Roche lobe radius will be $R_{\mathrm{L,A}} \simeq 0.45 a_f \approx 100 R_{\odot}$, which it should reach in the asymptotic giant branch (AGB) phase. Mass transfer will likely be dynamically unstable given that the envelope is convective and the mass ratio accretor to donor is relatively low \citep[e.g.][]{Hjellming1987}. At this point the primary has a helium core mass $M_{\mathrm{A,core}} \simeq 0.8 M_{\odot}$ and an envelope mass $M_{\mathrm{A,env}} \simeq 3 M_{\odot}$. The binding energy of the primary envelope is roughly 

\begin{align}
|E_{\mathrm{bind,env}}| &\simeq \frac{G M_{A} M_{\mathrm{A,env}}}{\lambda R_{A}} \\ &\simeq 4.8 \times 10^{47} \text{ erg} \left ( \frac{\lambda}{0.9} \right )^{-1} 
\end{align}

\noindent where $\lambda$ is a parameter that takes into account the structure of the envelope. From Table 1 in \cite{Dewis2000}, for a $M=4.0 M_{\odot}$ star expanded to $R \sim 100 R_{\odot}$, $\lambda = \lambda_g \sim 0.43$ when only including the gravitational binding energy and $\lambda = \lambda_b \sim 0.91$ when including all the thermal energy of the envelope as well.

Meanwhile, the orbital energy available before the core of the primary would merge with the secondary is 

\begin{align}
|\Delta E_{\mathrm{orb}}| &= \frac{G M_{B} M_{\mathrm{A,core}}}{2 a_f} - \frac{G M_{B} M_{A}}{2 a_i} \\ &\simeq 1.4 \times 10^{48} \text{ erg}
\end{align}

\noindent where $a_f \sim R_{B} \simeq 1.7 R_{\odot}$ is the final separation (B will barely evolve over the next 50 Myr that it will take for the primary to reach the AGB) and $a_i \simeq 1.05 \text{ au}$ is the initial separation at the common envelope phase. 

It is therefore likely that the system will be able to expel the common envelope and become a close white dwarf plus main sequence binary that will eventually become a symbiotic star once the companion also expands on its much longer nuclear timescale. 

\section{Conclusion}
\label{sec:conclusion}

The four early-type stars that make up the Teapot asterism are all multiple systems. Here we combined interferometric, spectroscopic, photometric and astrometric data to solve for the orbital and stellar parameters of the binary Nandou=Phi Sagittarii. It is a relatively standard binary, with a $M_B \simeq 1.6 M_{\odot}$ main-sequence secondary and a $M_A \simeq 4 M_{\odot}$ primary that is most likely at the end of its main-sequence phase. The system will probably become a white dwarf plus main sequence binary following a common envelope ejection when the primary reaches the AGB and eventually a symbiotic star once the secondary also expands towards becoming a red giant. 

We note two important takeaways from our analysis. The first is that the radial velocity measurements from very old papers are accurate and have a precision of about $3 \text {km}\text{ s}^{-1}$. The second is that there should be a great deal of care in interpreting photocenter semi-major axes using IAD from the Hipparcos 2007 re-reduction, as the more aggressive modeling seems to overfit and significantly suppress the orbital motion signature. The IAD from the Hipparcos original reduction seem to be much more immune to this at the expense of larger errorbars. 

It will be interesting to see whether the upcoming Gaia DR4 will be able to solve for the photocenter orbit of Phi Sagittarii. If so, a more precise parallax should lead to much better constrained dynamical masses. 

\section*{Acknowledgments}

This research has made use of the Jean-Marie Mariotti Center \texttt{SearchCal} service \footnote{Available at http://www.jmmc.fr/searchcal} co-developped by LAGRANGE and IPAG, CDS Astronomical Databases SIMBAD and VIZIER \footnote{Available at http://cdsweb.u-strasbg.fr/}, NASA's Astrophysics Data System Bibliographic Services, NumPy \citep{van2011numpy} and matplotlib, a Python library for publication quality graphics \citep{Hunter2007}. 

\section*{Data availability}
The data underlying this article is publicly available from the respective archives. Reduced data can be provided by the authors on request. 

\bibliographystyle{aasjournal}
\bibliography{main}{}

\appendix

\section{Supplementary tables and figures}
\label{app}

\begin{figure*}[]
\centering
\includegraphics[width=\textwidth]{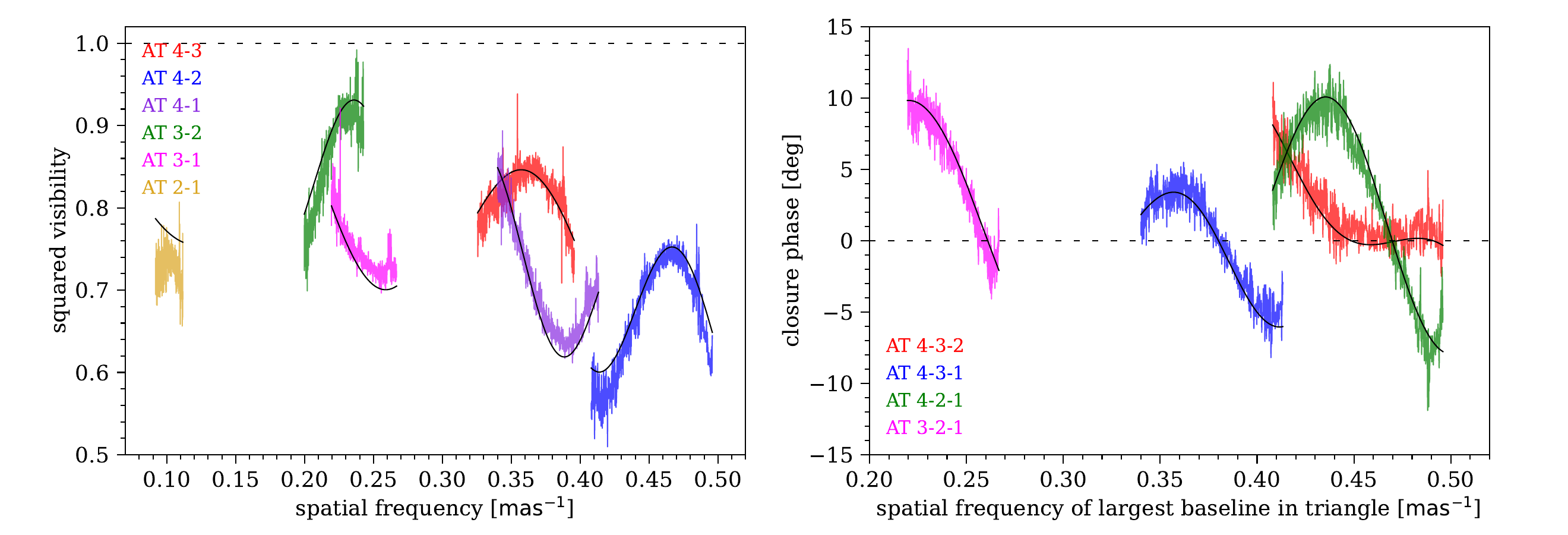} \\
\includegraphics[width=\textwidth]{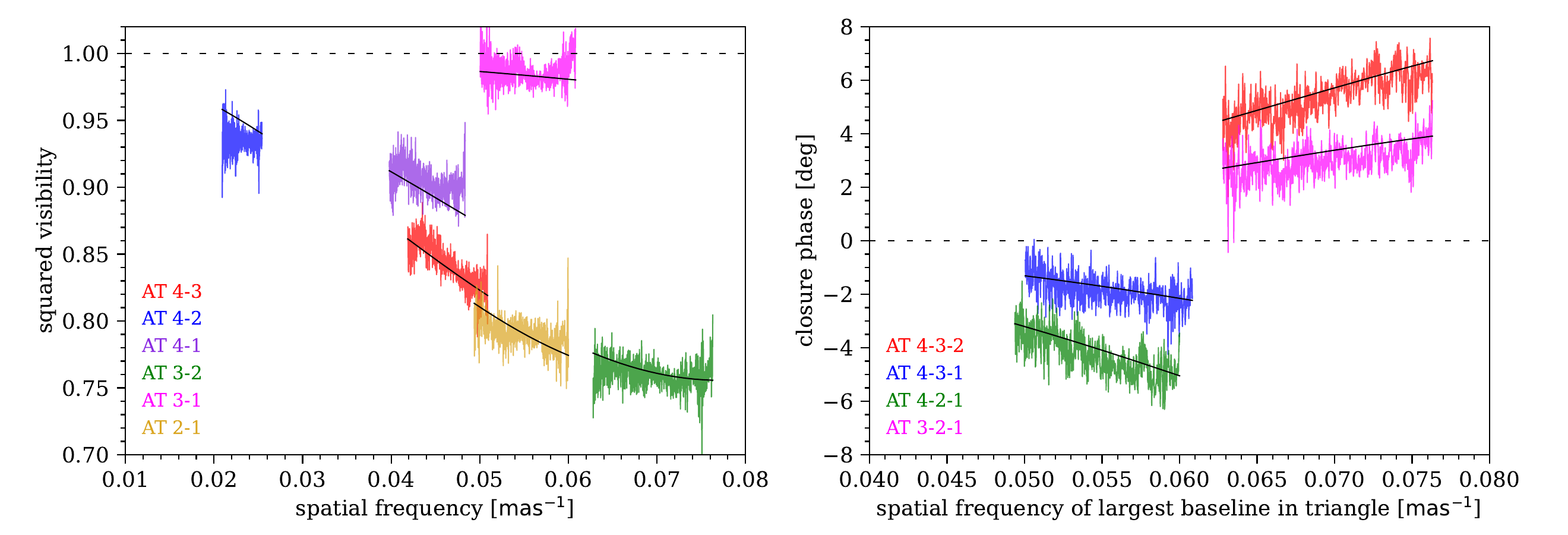} \\
\includegraphics[width=\textwidth]{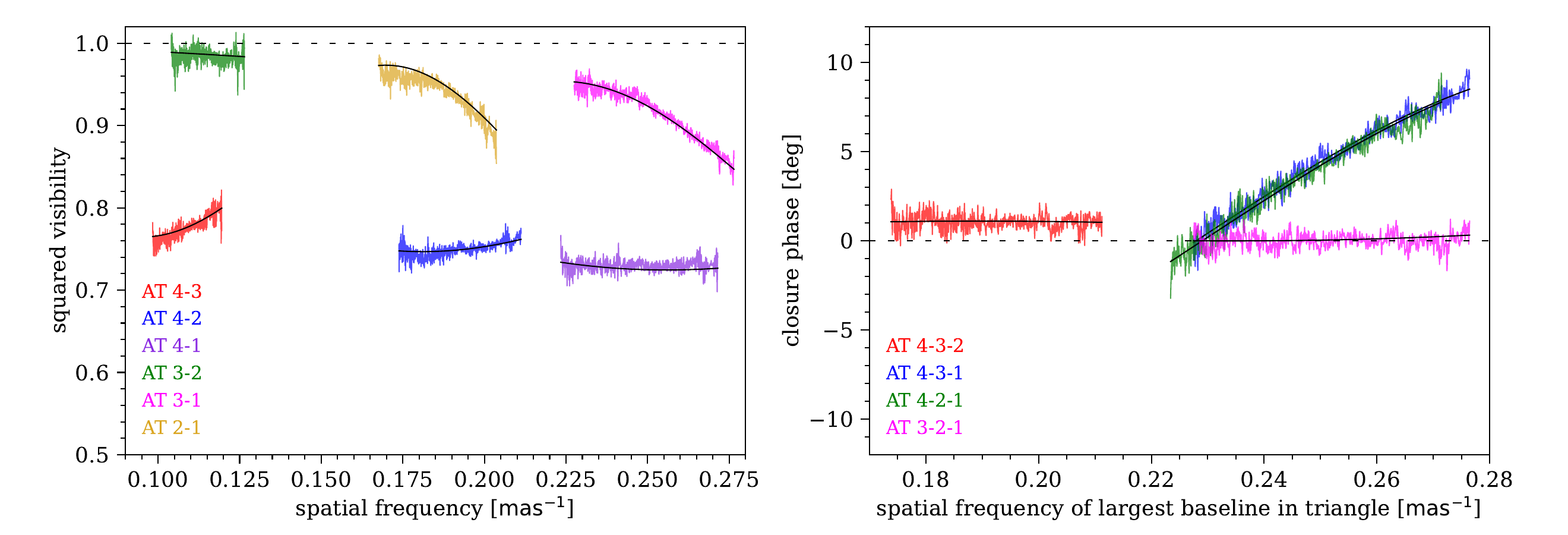} \\
\includegraphics[width=\textwidth]{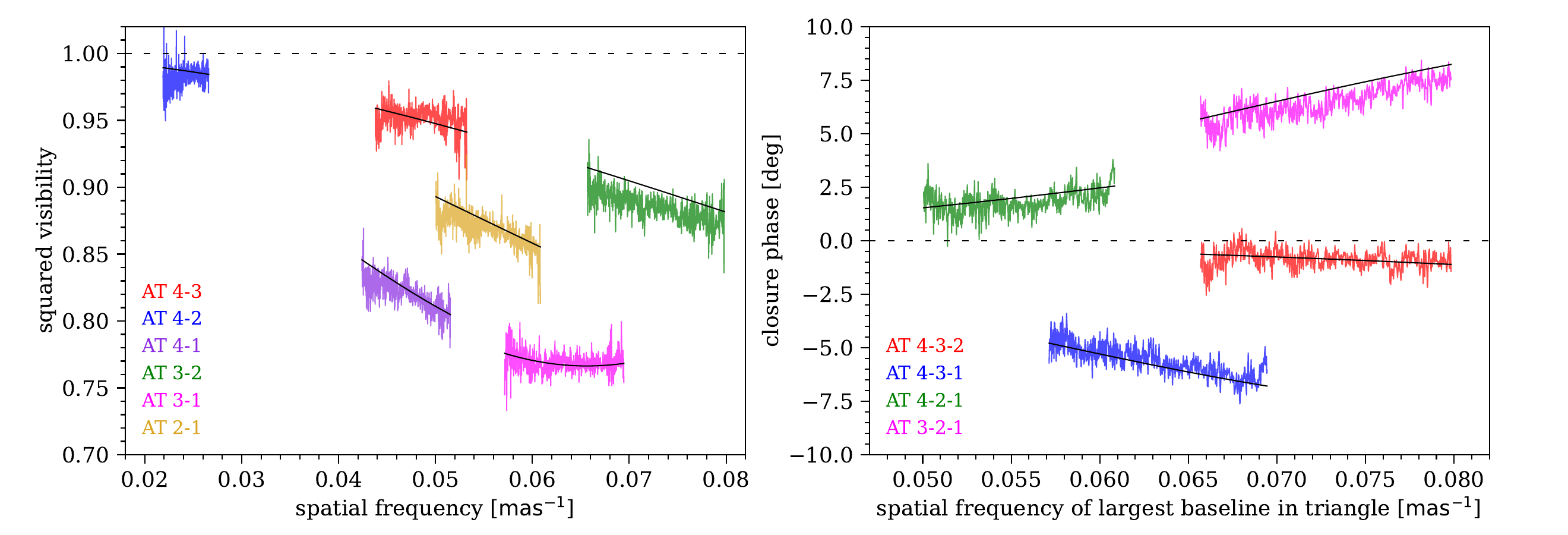}
\caption{\label{fig:gravity_fit_all} VLTI/GRAVITY data for Phi Sagittarii (colored) and best fit binary model (solid black) for epochs 2024-06-09, 2024-06-19, 2024-07-04 and 2024-07-26 (top to bottom).}
\end{figure*}

Figure \ref{fig:gravity_fit_all} shows the interferometric data and best fit binary model for the four remaining epochs. 

Table \ref{table:past_rv} lists the past radial velocity measurements of Phi Sgr. 

\begin{figure*}[]
\centering
\includegraphics[width=\textwidth]{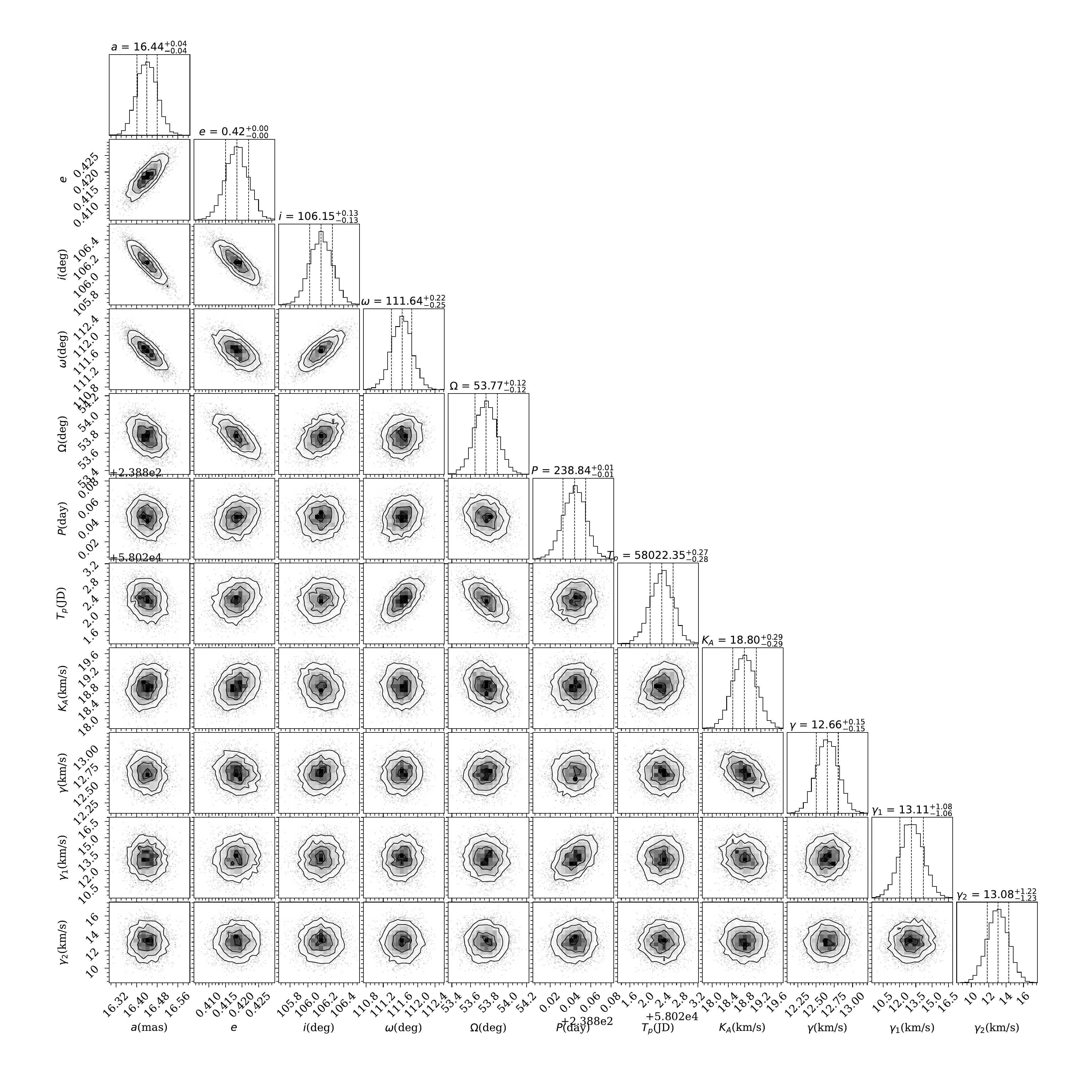}
\caption{\label{fig:orbital_fit_corner} Full orbital parameters distributions for Phi Sagittarii.}
\end{figure*}

Figure \ref{fig:orbital_fit_corner} shows the full distributions for orbital parameters of Phi Sgr based on the best fit parameters over 5000 resamples of the data. 

\begin{figure*}[]
\includegraphics[width=\columnwidth]{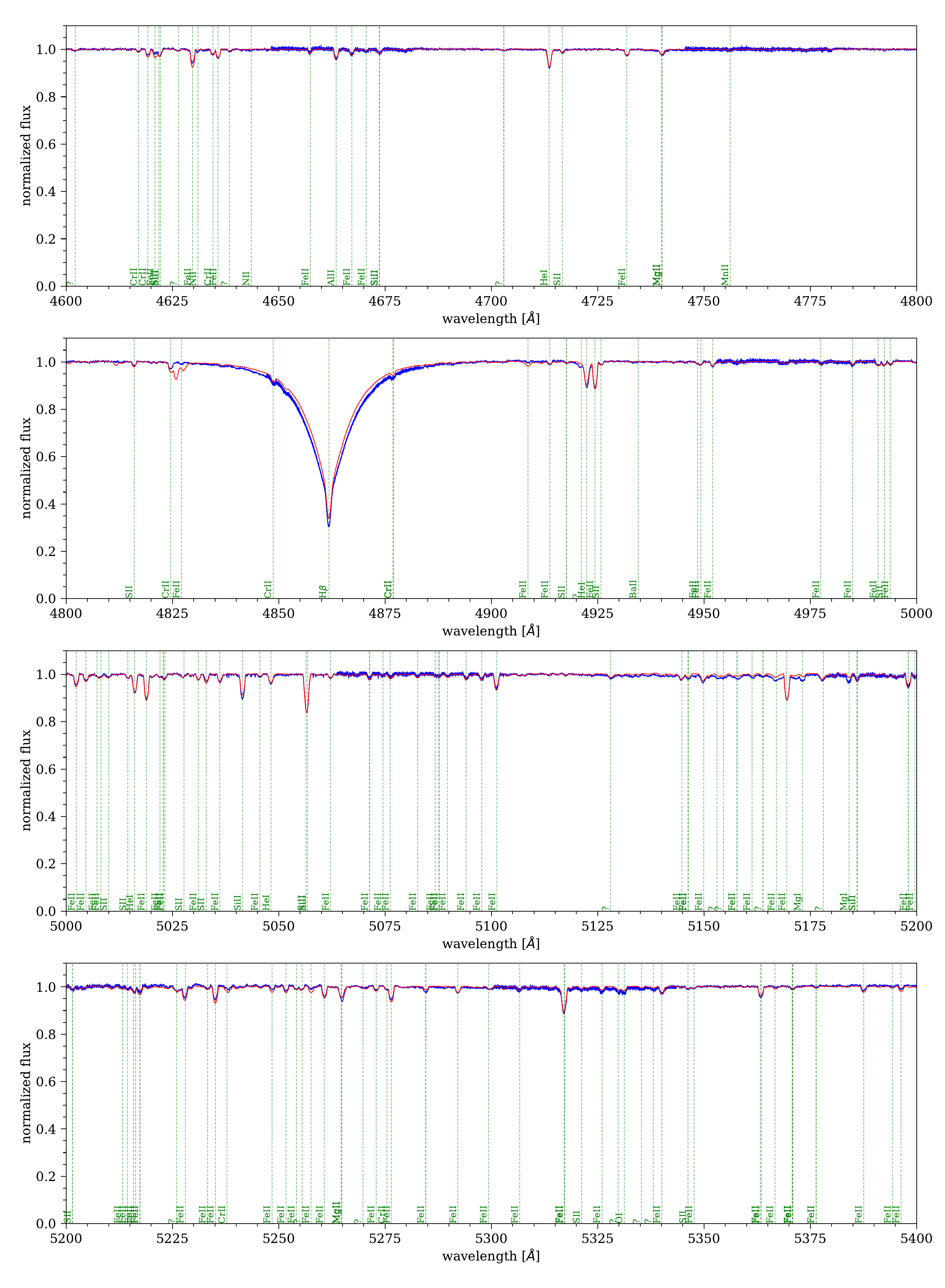}
\caption{ESPaDOns normalized spectrum of Phi Sagittarii (blue) for epoch 2015-04-03 and the best fit normalized BOSZ spectrum (red).}
\label{fig:spectrum_2}
\end{figure*}

\begin{figure*}[]
\includegraphics[width=\columnwidth]{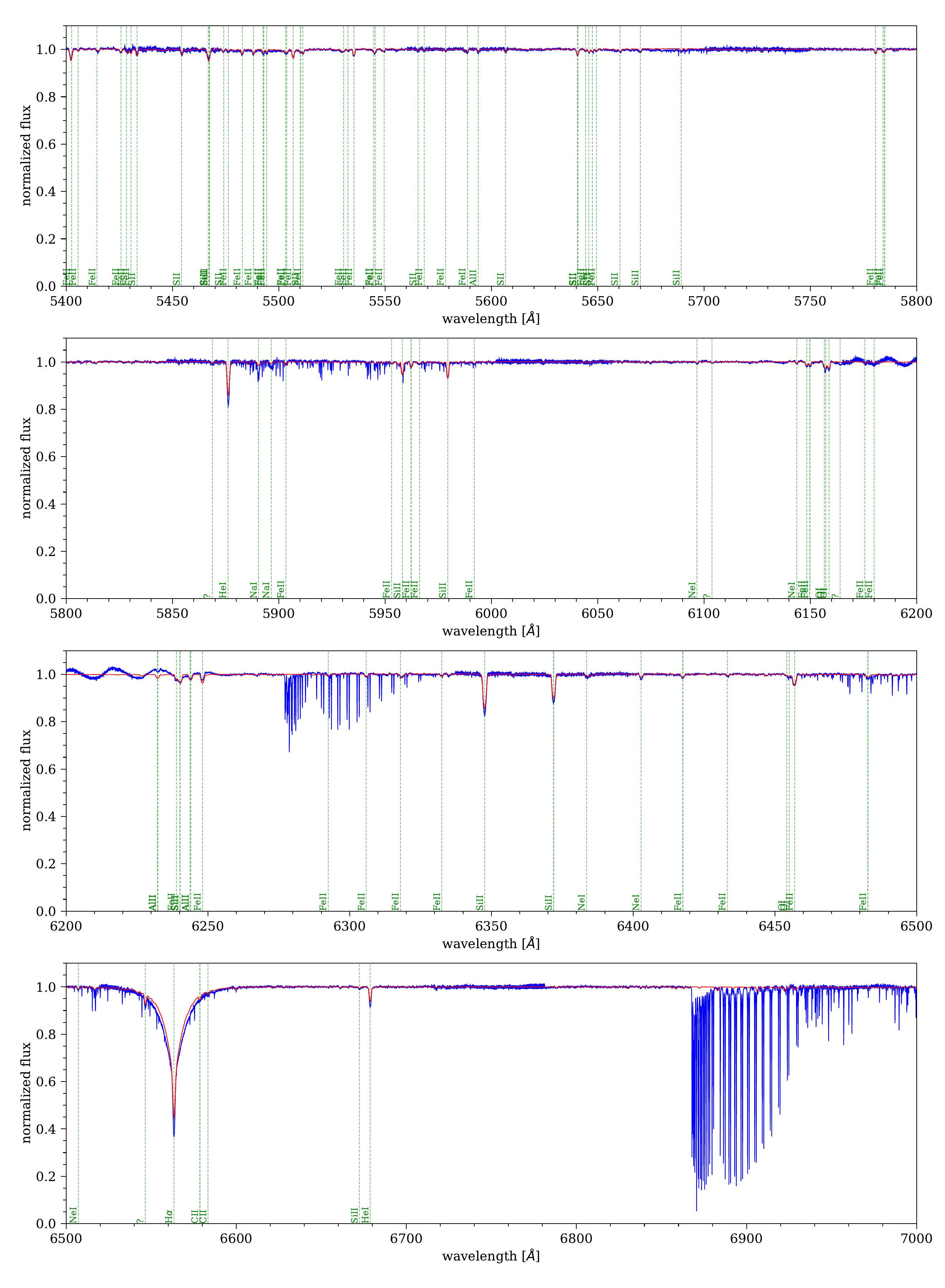}
\caption{ESPaDOns normalized spectrum of Phi Sagittarii (blue) for epoch 2015-04-03 and the best fit normalized BOSZ spectrum (red).}
\label{fig:spectrum_3}
\end{figure*}

\begin{figure*}[]
\includegraphics[width=\columnwidth]{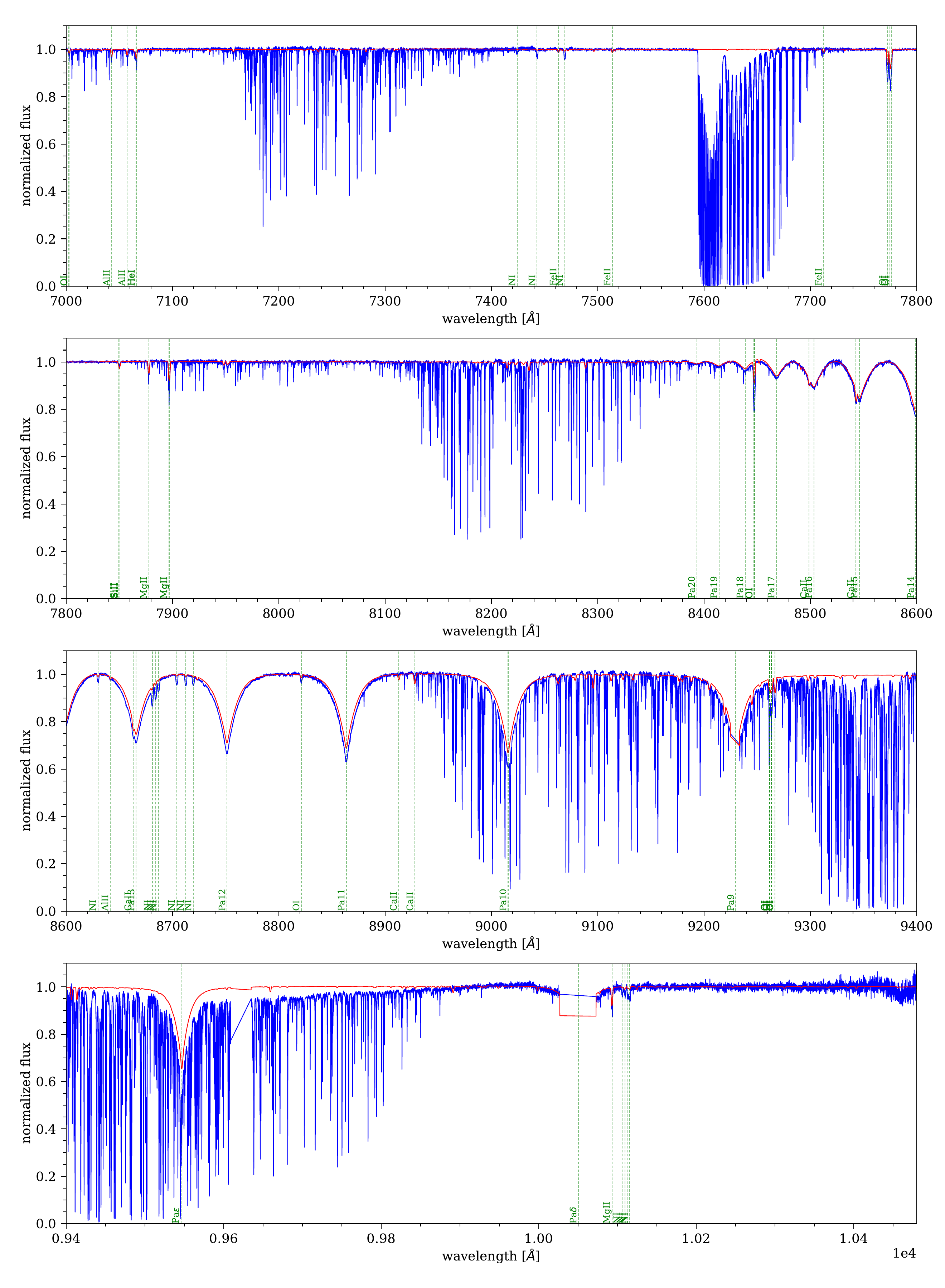}
\caption{ESPaDOns normalized spectrum of Phi Sagittarii (blue) for epoch 2015-04-03 and the best fit normalized BOSZ spectrum (red).}
\label{fig:spectrum_4}
\end{figure*}

Figures \ref{fig:spectrum_2}, \ref{fig:spectrum_3}, \ref{fig:spectrum_4} show the remaining panels of the ESPaDOns normalized spectrum for epoch 2015-04-03 and the corresponding best fit normalized BOSZ model.  

Table \ref{table:line_list} lists the spectral lines identified in the spectrum of Phi Sagittarii. 

\begin{table}
\centering
\caption{\label{table:past_rv} Past radial velocity measurements of Phi Sagittarii.}
\begin{tabular}{cc}
\hline \hline
JD & $v_{\mathrm{A}}$ \\ [0.3cm]

2416272.347 & $+20.8$ \\ [0.3cm]

2416643.283 & $-2.6$ \\ [0.3cm]

2418171.159 & $+38.0$ \\ [0.3cm]

2418175.079  & $+30.5$ \\ [0.3cm]

2418178.088 & $+26.1$ \\ [0.3cm]

2418179.105 & $+32.3$ \\ [0.3cm]

2418182.117 & $+35.0$ \\ [0.3cm]

2418444.351 & $+17.8$ \\ [0.3cm]

2418872.193 & $+33.3$ \\ [0.3cm]

2420649.306 & $+8.2$ \\ [0.3cm]

2416694.65 & $+14$ \\ [0.3cm]

2416981.02 & $+24$ \\ [0.3cm]

2417704.98 & $+30$ \\ [0.3cm]

2417748.69 & $+18$ \\ [0.3cm]

2418192.68  & $+25$ \\ [0.3cm]

2418521.66 & $+4$ \\ [0.3cm]

\hline
\end{tabular}
\tablenotetext{0}{Notes:}
\tablenotetext{0}{Sources: \cite{Lee1910,Frost1926,Campbell1928}.}
\end{table}

\clearpage

\renewcommand{\arraystretch}{1.4}
\begin{longtable}{cccccccc}
\caption{Identified spectral lines.} \label{table:line_list} \\
\toprule
Ion & Wavelength (\AA) & Ion & Wavelength (\AA) & Ion & Wavelength (\AA) & Ion & Wavelength (\AA) \\
\midrule
\endfirsthead
\toprule
Ion & Wavelength (\AA) & Ion & Wavelength (\AA) & Ion & Wavelength (\AA) & Ion & Wavelength (\AA) \\
\midrule
\endhead
\bottomrule
\endfoot
\midrule
H17 & 3697.15 & H16 & 3703.85 & H15 & 3711.97 & H14 & 3721.94 \\[2pt]
H13 & 3734.37 & H12 & 3750.15 & H11 & 3770.63 & H10 & 3797.90 \\[2pt]
H9 & 3835.38 & H8 & 3889.05 & H$\epsilon$ & 3970.07 & H$\delta$ & 4101.735 \\[2pt]
H$\gamma$ & 4340.463 & H$\beta$ & 4861.325 & H$\alpha$ & 6562.80 & Pa20 & 8392.45 \\[2pt]
Pa19 & 8413.38 & Pa18 & 8438.02 & Pa17 & 8467.31 & Pa16 & 8502.54 \\[2pt]
Pa15 & 8545.44 & Pa14 & 8598.46 & Pa13 & 8665.08 & Pa12 & 8750.54 \\[2pt]
Pa11 & 8862.85 & Pa10 & 9014.9 & Pa9 & 9229.0 & Pa$\epsilon$ & 9545.0 \\[2pt]
Pa$\delta$ & 10049.4 & & & & & & \\[2pt]
\midrule 
HeI & 3819.6074 & HeI & 3867.48 & HeI & 4009.256516 & HeI & 4026.1914 \\[2pt]
HeI & 4120.8154 & HeI & 4143.761 & HeI & 4168.967 & HeI & 4387.9296 \\[2pt]
HeI & 4437.551 & HeI & 4471.4802 & HeI & 4713.1457 & HeI & 4921.9313 \\[2pt]
HeI & 5015.6783 & HeI & 5047.73857 & HeI & 5875.621 & HeI & 6678.151 \\[2pt]
HeI & 7065.190 & HeI & 7065.71 \\[2pt]
\midrule
CII & 3918.978 & CII & 3920.693 & CII & 4267.003 & CII & 4267.258 \\[2pt]
CII & 6578.0481 & CII & 6582.8764 &  &  &  &  \\[2pt]
\midrule
NI & 7423.64 & NI & 7442.29 & NI & 7468.31 & NI & 8629.24 \\[2pt]
NI & 8680.28 & NI & 8683.40 & NI & 8686.15 & NI & 8703.25 \\[2pt]
NI & 8711.70 & NI & 8718.83 & NI & 10105.13 & NI & 10108.89 \\[2pt]
NI & 10112.48 & NI & 10114.64 &  &  &  &  \\[2pt]
\midrule
NII & 3995.00 & NII & 4630.54 & NII & 4643.08 &  &  \\[2pt]
\midrule
OI & 5330.74 & OI & 6155.98 & OI & 6156.77 & OI & 6158.18 \\[2pt]
OI & 6453.60 & OI & 6454.44 & OI & 7001.92 & OI & 7002.23 \\[2pt]
OI & 7771.94 & OI & 7774.17 & OI & 7775.39 & OI & 8446.25 \\[2pt]
OI & 8446.36 & OI & 8446.76 & OI & 8820.43 & OI & 9260.81 \\[2pt]
OI & 9260.84 & OI & 9260.94 & OI & 9262.58 & OI & 9262.67 \\[2pt]
OI & 9262.77 & OI & 9265.826 & OI & 9265.94 & OI & 9266.01 \\[2pt]
\midrule
NeI & 6096.1630 & NeI & 6143.0627 & NeI & 6382.9914 & NeI & 6402.248 \\[2pt]
NeI & 6506.5277 &  &  &  &  &  &  \\[2pt]
\midrule
NaI & 5889.95095 & NaI & 5895.92424 &  &  &  &  \\[2pt]
\midrule
MgI & 5172.6843 & MgI & 5183.6042 &  &  &  &  \\[2pt]
\midrule
MgII & 4384.637 & MgII & 4390.564 & MgII & 4427.994 & MgII & 4433.990 \\[2pt]
MgII & 4436.486 & MgII & 4436.598 & MgII & 4481.130 & MgII & 4481.150 \\[2pt]
MgII & 4481.327 & MgII & 4739.588 & MgII & 4739.712 & MgII & 5264.215 \\[2pt]
MgII & 5264.368 & MgII & 7877.051 & MgII & 7896.04 & MgII & 7896.368 \\[2pt]
MgII & 10092.160 &  &  &  &  &  &  \\[2pt]
\midrule
AlI & 3944.0058 & AlI & 3961.5200 &  &  &  &  \\[2pt]
\midrule
AlII & 4663.056 & AlII & 5593.302 & AlII & 6231.62 & AlII & 6231.745 \\[2pt]
AlII & 6243.20 & AlII & 6243.36 & AlII & 7042.06 & AlII & 7056.60 \\[2pt]
AlII & 8640.705 &  &  &  &  &  &  \\[2pt]
\midrule
SiII & 3853.66 & SiII & 3856.02 & SiII & 3862.60 & SiII & 4075.45 \\[2pt]
SiII & 4076.78 & SiII & 4128.07 & SiII & 4130.89 & SiII & 4200.658 \\[2pt]
SiII & 4200.887 & SiII & 4200.898 & SiII & 4621.42 & SiII & 4621.722 \\[2pt]
SiII & 4673.256 & SiII & 4673.284 & SiII & 5041.03 & SiII & 5055.98 \\[2pt]
SiII & 5056.317 & SiII & 5185.520 & SiII & 5185.555 & SiII & 5466.43 \\[2pt]
SiII & 5466.894 & SiII & 5669.56 & SiII & 5688.81 & SiII & 5957.56 \\[2pt]
SiII & 5978.93 & SiII & 6239.61 & SiII & 6239.66 & SiII & 6347.10 \\[2pt]
SiII & 6371.36 & SiII & 6671.88 & SiII & 7848.80 & SiII & 7849.72 \\[2pt]
\midrule
SiIII & 4552.62 & SiIII & 4567.82 &  &  &  &  \\[2pt]
\midrule
SII & 4032.779 & SII & 4153.064 & SII & 4162.665 & SII & 4278.500 \\[2pt]
SII & 4282.595 & SII & 4294.398 & SII & 4463.579 & SII & 4464.430 \\[2pt]
SII & 4483.429 & SII & 4524.947 & SII & 4716.267 & SII & 4815.549 \\[2pt]
SII & 4917.212 & SII & 4925.347 & SII & 4991.974 & SII & 5009.564 \\[2pt]
SII & 5014.069 & SII & 5027.221 & SII & 5032.447 & SII & 5201.024 \\[2pt]
SII & 5320.732 & SII & 5345.721 & SII & 5432.815 & SII & 5453.828 \\[2pt]
SII & 5473.620 & SII & 5509.718 & SII & 5564.976 & SII & 5606.151 \\[2pt]
SII & 5639.972 & SII & 5640.333 & SII & 5647.033 & SII & 5659.985 \\[2pt]
\midrule
CaII & 3933.66 & CaII & 3968.47 & CaII & 8498.02 & CaII & 8542.09 \\[2pt]
CaII & 8662.14 & CaII & 8912.07 & CaII & 8927.36 &  &  \\[2pt]
\midrule
TiII & 3759.2915 & TiII & 3761.3202 & TiII & 3900.5389 & TiII & 3913.4614 \\[2pt]
TiII & 4053.8210 & TiII & 4163.6437 & TiII & 4290.2148 & TiII & 4300.0421 \\[2pt]
TiII & 4307.8657 & TiII & 4312.8600 & TiII & 4314.9708 & TiII & 4395.0312 \\[2pt]
TiII & 4443.8007 & TiII & 4468.4924 & TiII & 4501.2699 & TiII & 4533.9600 \\[2pt]
TiII & 4549.6216 & TiII & 4563.7575 & TiII & 4571.9713 &  &  \\[2pt]
\midrule
CrII & 4242.36595 & CrII & 4261.91683 & CrII & 4275.56085 & CrII & 4284.1904 \\[2pt]
CrII & 4554.98902 & CrII & 4558.64422 & CrII & 4588.19836 & CrII & 4592.05241 \\[2pt]
CrII & 4616.62424 & CrII & 4618.80671 & CrII & 4634.07286 & CrII & 4824.13069 \\[2pt]
CrII & 4848.24690 & CrII & 4876.39731 & CrII & 4876.4891 & CrII & 5237.32185 \\[2pt]
CrII & 5274.97462 &  &  &  &  &  &  \\[2pt]
\midrule
MnII & 4755.733 &  &  &  &  &  &  \\[2pt]
\midrule
FeII & 3759.460 & FeII & 3814.121 & FeII & 3824.913 & FeII & 3827.079 \\[2pt]
FeII & 3906.037 & FeII & 3914.480 & FeII & 3935.942 & FeII & 3938.289 \\[2pt]
FeII & 3938.969 & FeII & 4024.552 & FeII & 4122.638 & FeII & 4173.450 \\[2pt]
FeII & 4178.8546 & FeII & 4233.1627 & FeII & 4250.429 & FeII & 4258.155 \\[2pt]
FeII & 4273.317 & FeII & 4296.567 & FeII & 4303.168 & FeII & 4351.7629 \\[2pt]
FeII & 4357.571 & FeII & 4369.400 & FeII & 4384.31313 & FeII & 4385.379 \\[2pt]
FeII & 4402.879 & FeII & 4416.818 & FeII & 4451.542 & FeII & 4461.7108 \\[2pt]
FeII & 4489.185 & FeII & 4491.400 & FeII & 4493.525 & FeII & 4508.2805 \\[2pt]
FeII & 4515.3344 & FeII & 4520.2214 & FeII & 4522.6278 & FeII & 4541.518 \\[2pt]
FeII & 4549.197 & FeII & 4549.4665 & FeII & 4555.8881 & FeII & 4576.328 \\[2pt]
FeII & 4582.835 & FeII & 4583.8292 & FeII & 4596.0084 & FeII & 4598.4853 \\[2pt]
FeII & 4620.513 & FeII & 4629.3317 & FeII & 4635.3177 & FeII & 4656.981 \\[2pt]
FeII & 4666.744 & FeII & 4670.170 & FeII & 4731.439 & FeII & 4826.6825 \\[2pt]
FeII & 4908.1454 & FeII & 4913.2951 & FeII & 4923.9216 & FeII & 4948.0872 \\[2pt]
FeII & 4948.7913 & FeII & 4951.5805 & FeII & 4977.0332 & FeII & 4984.4852 \\[2pt]
FeII & 4990.5062 & FeII & 4993.355 & FeII & 5001.9529 & FeII & 5004.1881 \\[2pt]
FeII & 5006.8402 & FeII & 5007.7338 & FeII & 5018.4354 & FeII & 5021.5907 \\[2pt]
FeII & 5022.4183 & FeII & 5022.7888 & FeII & 5030.6317 & FeII & 5035.6999 \\[2pt]
FeII & 5045.1067 & FeII & 5061.7101 & FeII & 5070.8959 & FeII & 5074.051 \\[2pt]
FeII & 5075.7599 & FeII & 5082.2264 & FeII & 5086.3071 & FeII & 5087.2222 \\[2pt]
FeII & 5087.3000 & FeII & 5089.2129 & FeII & 5093.5719 & FeII & 5097.2737 \\[2pt]
FeII & 5100.854 & FeII & 5144.3515 & FeII & 5145.7299 & FeII & 5145.8220 \\[2pt]
FeII & 5149.4624 & FeII & 5157.2834 & FeII & 5157.2834 & FeII & 5160.840 \\[2pt]
FeII & 5166.5511 & FeII & 5169.0282 & FeII & 5197.5678 & FeII & 5199.1186 \\[2pt]
FeII & 5212.8315 & FeII & 5213.9516 & FeII & 5215.3470 & FeII & 5215.8459 \\[2pt]
FeII & 5216.8585 & FeII & 5216.8585 & FeII & 5227.582 & FeII & 5232.7809 \\[2pt]
FeII & 5234.6236 & FeII & 5247.9553 & FeII & 5251.2252 & FeII & 5253.6484 \\[2pt]
FeII & 5257.1189 & FeII & 5260.2544 & FeII & 5272.413 & FeII & 5275.9968 \\[2pt]
FeII & 5284.092 & FeII & 5291.6609 & FeII & 5298.8574 & FeII & 5306.1820 \\[2pt]
FeII & 5316.6098 & FeII & 5316.777 & FeII & 5325.543 & FeII & 5339.5943 \\[2pt]
FeII & 5347.1873 & FeII & 5362.7509 & FeII & 5362.9698 & FeII & 5366.2087 \\[2pt]
FeII & 5370.2772 & FeII & 5370.3163 & FeII & 5370.4338 & FeII & 5375.8423 \\[2pt]
FeII & 5375.8423 & FeII & 5387.0633 & FeII & 5393.8401 & FeII & 5395.8546 \\[2pt]
FeII & 5402.0576 & FeII & 5405.0921 & FeII & 5414.089 & FeII & 5425.269 \\[2pt]
FeII & 5427.818 & FeII & 5429.9873 & FeII & 5466.912 & FeII & 5475.8260 \\[2pt]
FeII & 5482.3051 & FeII & 5487.6181 & FeII & 5492.0782 & FeII & 5492.3979 \\[2pt]
FeII & 5493.8306 & FeII & 5502.6698 & FeII & 5503.2150 & FeII & 5506.1985 \\[2pt]
FeII & 5510.7831 & FeII & 5529.937 & FeII & 5532.0859 & FeII & 5534.839 \\[2pt]
FeII & 5544.2006 & FeII & 5544.7580 & FeII & 5549.0005 & FeII & 5567.835 \\[2pt]
FeII & 5577.9121 & FeII & 5588.2212 & FeII & 5643.8695 & FeII & 5645.3901 \\[2pt]
FeII & 5648.8996 & FeII & 5780.1279 & FeII & 5783.6230 & FeII & 5784.4454 \\[2pt]
FeII & 5902.8238 & FeII & 5952.5183 & FeII & 5961.7058 & FeII & 5965.6199 \\[2pt]
FeII & 5991.3721 & FeII & 6147.734 & FeII & 6149.231 & FeII & 6175.135 \\[2pt]
FeII & 6179.378 & FeII & 6238.375 & FeII & 6247.559 & FeII & 6291.828 \\[2pt]
FeII & 6305.294 & FeII & 6317.374 & FeII & 6331.9551 & FeII & 6416.9303 \\[2pt]
FeII & 6432.6772 & FeII & 6456.3805 & FeII & 6482.2047 & FeII & 7462.414 \\[2pt]
FeII & 7513.1767 & FeII & 7711.7219 &  &  &  &  \\[2pt]
\midrule
FeIII & 4419.599 &  &  &  &  &  &  \\[2pt]
\midrule
SrII & 4077.714 & SrII & 4215.524 &  &  &  &  \\[2pt]
\midrule
BaII & 4554.033 & BaII & 4934.077 &  &  &  &  \\[2pt]
\end{longtable}

\end{document}